%% file: TEKIN_PREPOTS_ARXIV.tex
\documentclass[11pt]{scrartcl}

\makeatletter
    \DeclareOldFontCommand{\rm}{\normalfont\rmfamily}{\mathrm}
    \DeclareOldFontCommand{\sf}{\normalfont\sffamily}{\mathsf}
    \DeclareOldFontCommand{\tt}{\normalfont\ttfamily}{\mathtt}
    \DeclareOldFontCommand{\bf}{\normalfont\bfseries}{\mathbf}
    \DeclareOldFontCommand{\it}{\normalfont\itshape}{\mathit}
    \DeclareOldFontCommand{\sl}{\normalfont\slshape}{\@nomath\sl}
    \DeclareOldFontCommand{\sc}{\normalfont\scshape}{\@nomath\sc}
\makeatother

\RequirePackage[autooneside=false]{scrlayer-scrpage}
	\setkomafont{title}{\LARGE\sffamily\bfseries}
	\setkomafont{section}{\Large\sffamily\bfseries}
	\RedeclareSectionCommand[   afterskip=0.1cm,
	                            beforeskip=1cm]{section}

\usepackage{amsmath}
\usepackage{amssymb}
\usepackage{amstext}
\usepackage[auth-sc-lg]{authblk}
\usepackage[greek,english]{babel}
\usepackage{bm}
\usepackage{bold-extra}
\usepackage{color}
\usepackage[left=2.6cm,right=2.6cm,top=2.1cm,bottom=2.1cm,footskip=1cm,marginparwidth=2cm]{geometry}
\usepackage{xcolor}
\usepackage{graphicx}
\usepackage{import}
\usepackage{mathtools}
\usepackage{setspace}
\usepackage{slashed}
\usepackage{hyperref} 	
	\hypersetup{    colorlinks,
		            linkcolor={red!50!black},
		            citecolor={blue!50!black},
		            urlcolor={blue!80!black}, 
		            pdfstartview={XYZ null null 1},
		            bookmarksnumbered
		            }

\newcommand{\realF}{\mathcal{F}}
\newcommand{\M}{\mathcal{M}}
\newcommand{\ds}{\displaystyle}
\newcommand{\w}{\wedge}
\newcommand{\g}{\texttt{g}}
\newcommand{\GG}{\breve{\g}}
\newcommand{\h}{\texttt{h}}
\newcommand{\etaT}{\text{\large\begin{otherlanguage}{greek}\texttt{\texteta}\end{otherlanguage}}}
\newcommand{\etaTs}{\text{\begin{otherlanguage}{greek}\texttt{\texteta}\end{otherlanguage}}}
\newcommand{\psiT}{\text{\large\begin{otherlanguage}{greek}\texttt{\textxi}\end{otherlanguage}}}
\newcommand{\T}{\texttt{T}}
\newcommand{\wt}[1]{\widetilde{#1}}
\newcommand{\wh}[1]{\widehat{#1}}
\newcommand{\Lie}{\mathcal{L}}
\newcommand{\sLie}{\wh{\mathcal{L}}}
\newcommand{\tensor}{\otimes}
\newcommand{\QUAD}[1]{\quad\text{#1}\quad}
\newcommand{\QQUAD}[1]{\qquad\text{#1}\qquad}
\newcommand{\W}{\ell_{0}}
\newcommand{\NABLA}[1]{\nabla^{(#1)}}
\newcommand{\LAP}[1]{\text{Lap}^{(#1)}}
\newcommand{\DIV}[1]{\text{Div}^{(#1)}}
\newcommand{\EIN}[1]{\texttt{Ein}^{(#1)}}
\newcommand{\TR}[1]{\text{Tr}^{(#1)}}
\newcommand{\RIC}[1]{\texttt{Ric}^{(#1)}}

\newcommand{\TRREV}[1]{\mu^{(#1)}}

\newcommand{\CVRR}[1]{ \mathcal{R}^{\g^{(#1)}} }
\newcommand{\HMAX}{\mathcal{H}}
\newcommand{\taumax}{\tau_{\text{max}}}
\newcommand{\AR}{\mathcal{A}(\taumax)}

\newcommand{\CF}{\mathcal{C}}
\newcommand{\UU}{\mathcal{U}}
\newcommand{\UUP}{\mathcal{P}_{\mathcal{U}}}
\newcommand{\df}{\frac{}{}}
\newcommand{\PHYSDIM}[1]{\pmb{\left[\vphantom{#1}\right.} #1 \pmb{\left.\vphantom{#1}\right]}}

\newcommand{\proj}{\textrm{P}}
\newcommand{\PROJ}{\mathcal{P}}

\newcommand{\RR}{\mathbb{R}}
\newcommand{\CC}{\mathbb{C}}
\newcommand{\om}[2]{\omega^{#1}{}_{#2}}
\newcommand{\FORANY}{\text{for any}}
\newcommand{\FORALL}{\text{for all}}
\newcommand{\FM}{\mathcal{F}(\M)}
\newcommand{\RIEM}{\texttt{Riem}}
\newcommand{\inv}{\mbox{\tiny $-1$}}
\newcommand{\mbasis}[1]{\pmb{m}_{#1}}
\newcommand{\metscale}{\Upsilon}
\let\svthefootnote\thefootnote
	\newcommand{\freefootnote}[1]{%
		\let\thefootnote\relax%
		\footnotetext{#1}%
		\let\thefootnote\svthefootnote%
		}

\newcommand{\LASERSPACECURVES}[1]{%
	\begin{figure}[!ht]
		\centering
    \def\svgwidth{0.9\columnwidth}
    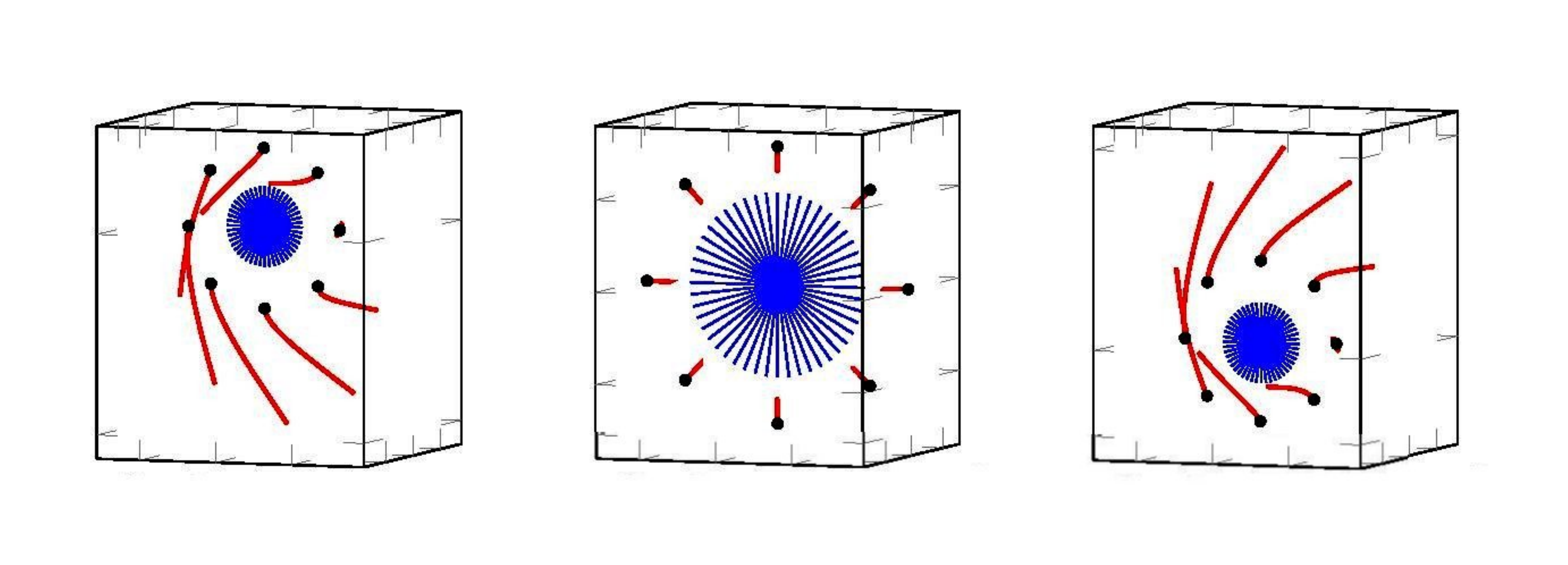
	
		\caption{#1}
		\label{fig:laserspacecurves}
	\end{figure} 
	}
\newcommand{\LASERCOMPACT}[1]{%
	\begin{figure}[!ht]
		\centering
    \def\svgwidth{0.6\columnwidth}
    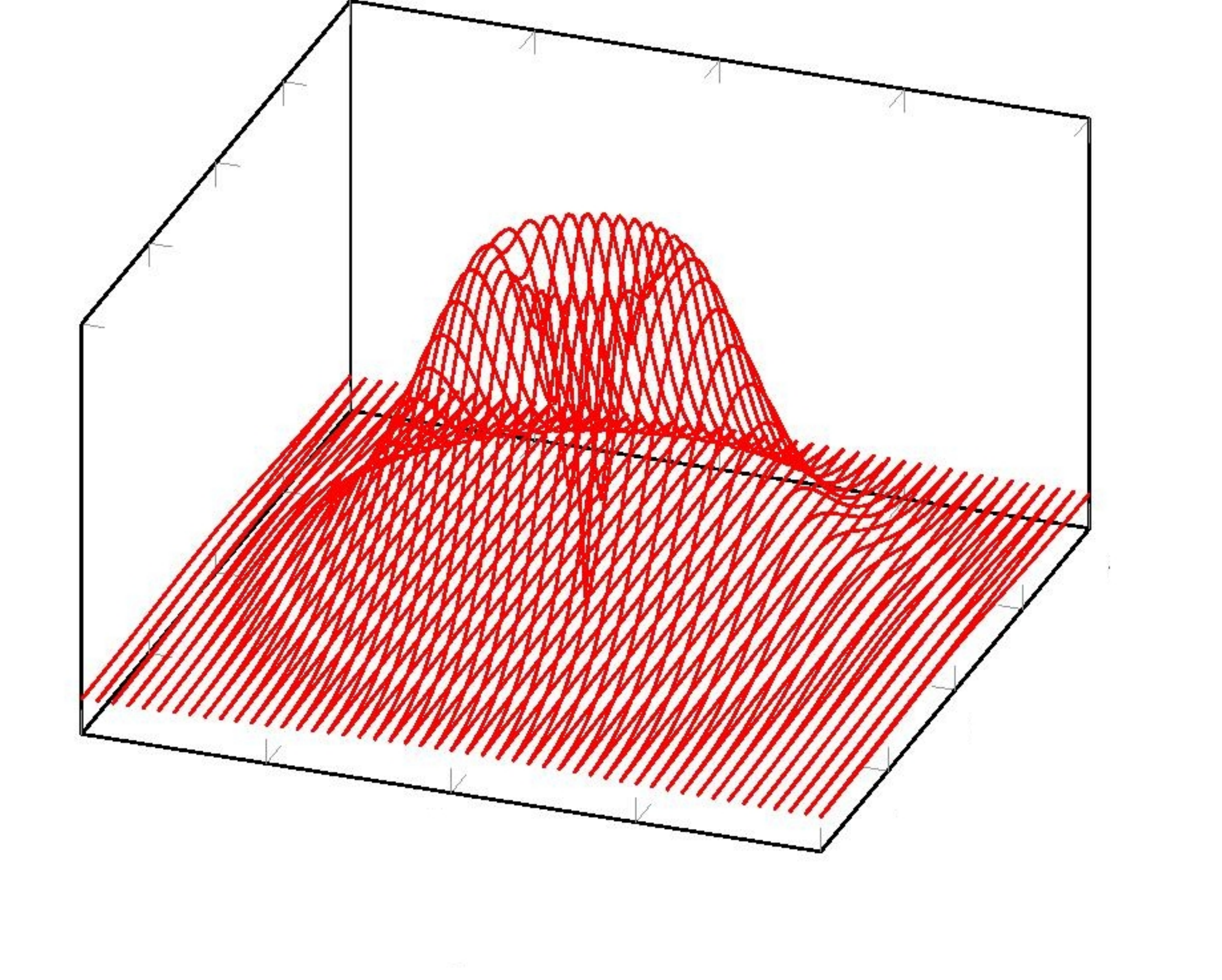
	
		\caption{#1}
		\label{fig:lasercompact}
	\end{figure} 
	}
\newcommand{\HMAXRRZP}[2]{%
	\begin{figure}[!ht]
		\centering
		\includegraphics[width=#1\textwidth]{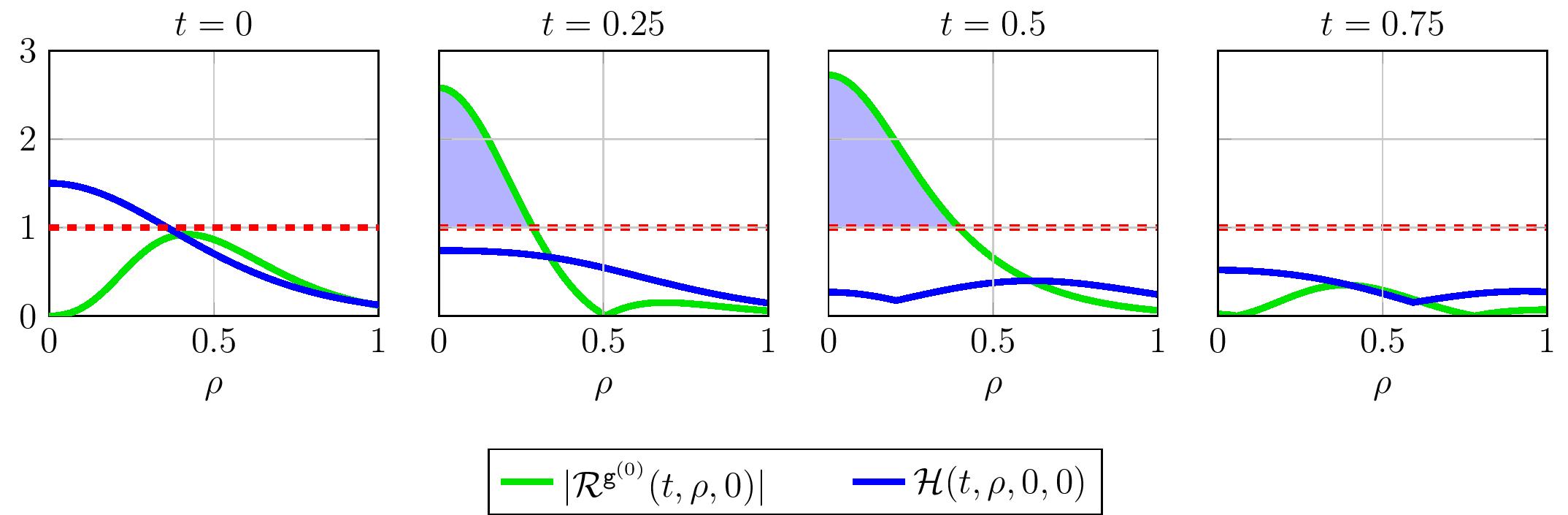}
		\caption{ #2 }
		\label{fig:hmaxrrzp}
	\end{figure}
	}
\newcommand{\RRZP}[2]{%
	\begin{figure}[!ht]
		\centering
		\includegraphics[width=#1\textwidth]{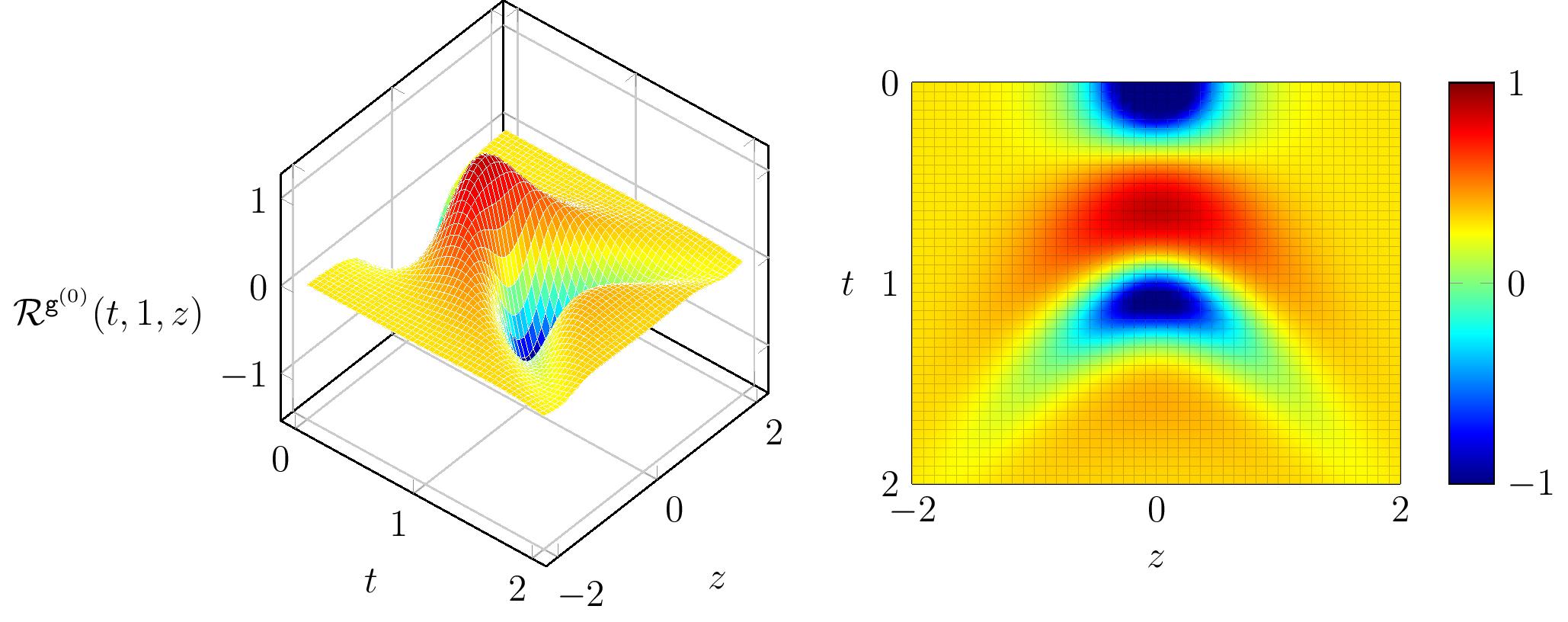}
		\caption{ #2 }
		\label{fig:rrzp}
	\end{figure}
	}
\newcommand{\ZPZLAYERS}[2]{%
	\begin{figure}[!ht]
		\centering
		\includegraphics[width=#1\textwidth]{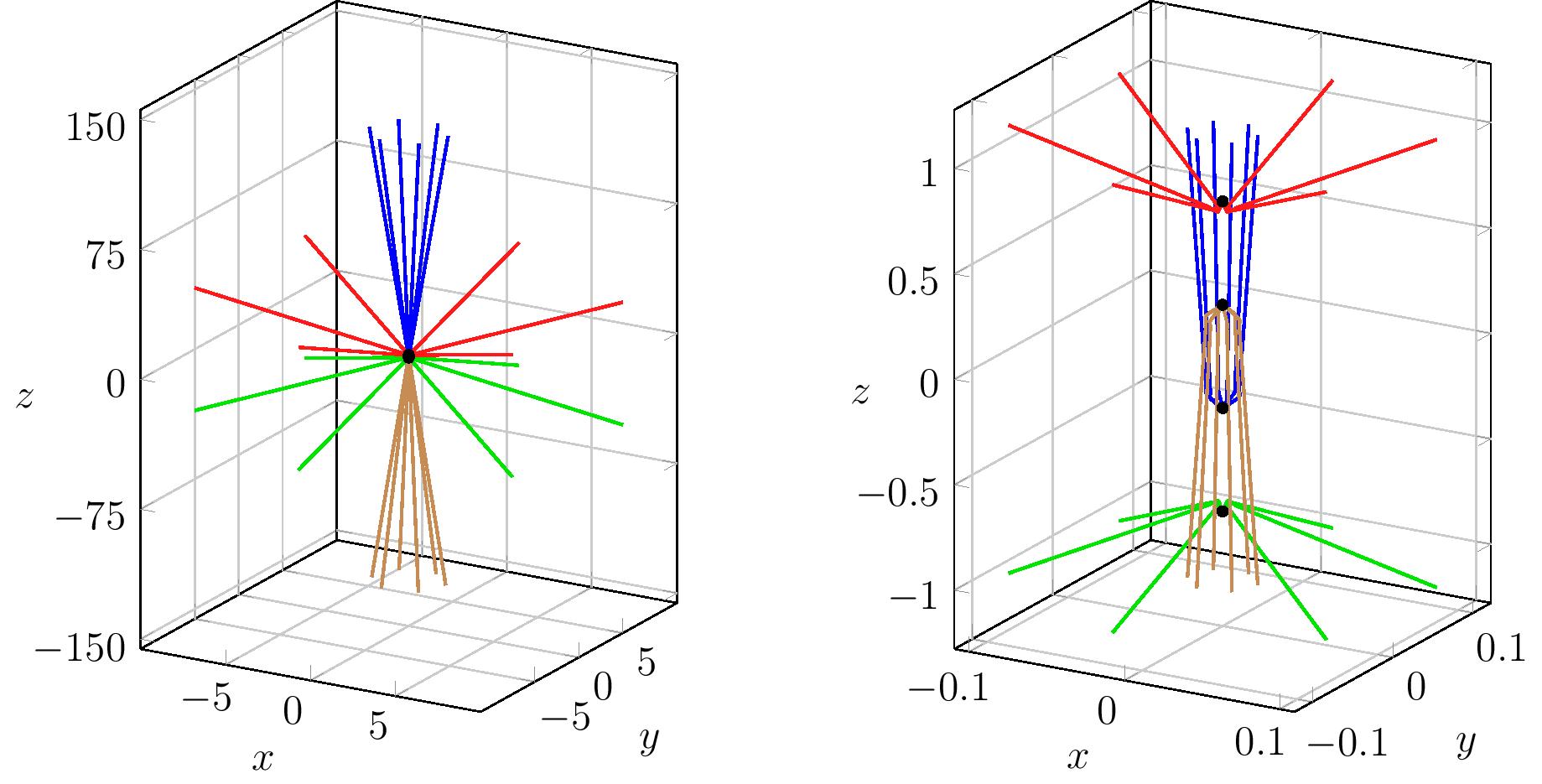}
		\caption{ #2 }
		\label{fig:zpzlayers}
	\end{figure}
	}
\newcommand{\ZPZLAYERSQ}[2]{%
	\begin{figure}[!ht]
		\centering
		\includegraphics[width=#1\textwidth]{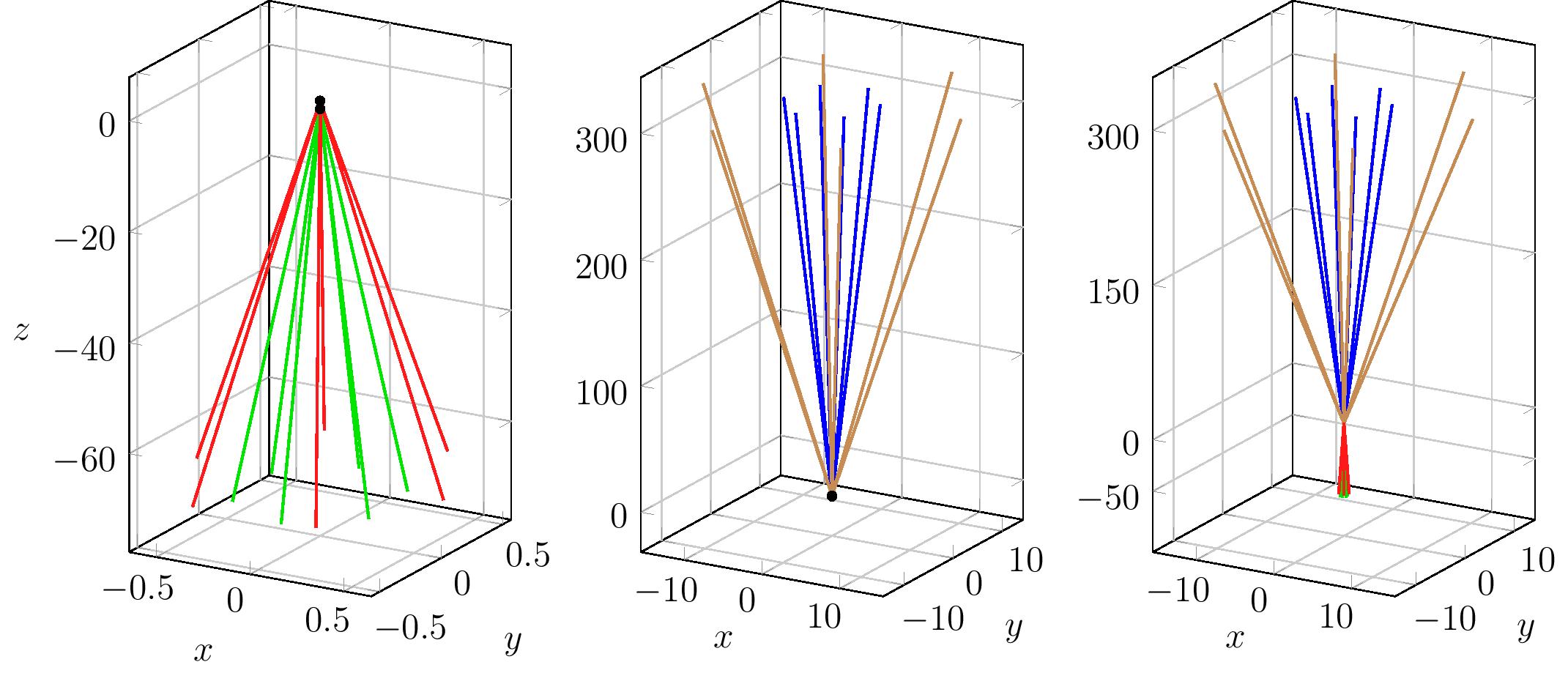}
		\caption{ #2 }
		\label{fig:zpzlayersq}
	\end{figure}
	}
	
\begin{document}
\title{Scalar Pre-potentials for Spinor and Tensor Fields on Spacetime}
\author{R. W. Tucker${}^{1}$ and T. J. Walton${}^{2,\mbox{\small$\dagger$}}$}
\date{}
\maketitle
\vspace{-1.5cm}		
\begin{center}
	${}^{1}$\parbox[t]{0.92\textwidth}{\small \raggedright %
		Department of Physics, University of Lancaster and Cockcroft Institute, Daresbury Laboratory, Warrington, UK
		} \\[0.3cm]
	${}^{2}$\parbox[t]{0.92\textwidth}{\small \raggedright %
		Jeremiah Horrocks Institute, School of Natural Sciences, University of Central Lancashire, Preston, UK
		} \\[0.4cm]
	E-mail: ${}^{1}$\href{mailto:r.tucker@lancaster.ac.uk}{r.tucker@lancaster.ac.uk} \;and\; ${}^{2}$\href{mailto:twalton@uclan.ac.uk}{twalton@uclan.ac.uk}
\end{center}
\begin{abstract}
	{\bf Abstract}. \quad We review a technique for solving a class of classical linear partial differential systems of relevance to physics in Minkowski spacetime. All the equations are amenable to analysis in terms of complex solutions in the kernel of the scalar Laplacian and a complexified Hertz potential. The complexification prescription ensures the existence of regular physical solutions with chirality and propagating, non-singular, pulse-like characteristics that are bounded in all three spatial dimensions. The technique is applied to the source-free Maxwell, Bopp-Land\'{e}-Podolsky and linearised Einstein field systems, and particular solutions are used for constructing classical models describing single-cycle laser pulses and a mechanism is discussed for initiating astrophysical jets. Our article concludes with a brief introduction to spacetime Clifford algebra ideals that we use to represent spinor fields. We employ these to demonstrate how the same technique used for tensor fields enables one to construct new propagating, chiral, non-singular, pulse-like spinor solutions to the massless Dirac equation in Minkowski spacetime. 
\end{abstract}
\freefootnote{${}^{\mbox{\small$\dagger$}}$Corresponding author.}

\section{Introduction}\label{sect:INTRO}
In this article we discuss a class of solutions to a set of linear field equations on Minkowski spacetime. The class of solutions will all be constructed from a scalar field $\alpha$ satisfying $\Box\,\alpha=0$, where $\Box$ is the Lorentz-invariant {\it D'Alembertian} (or {\it Laplacian operator}), and a set of covariantly-constant, antisymmetric tensor fields ($2$-forms) $\Pi^{\nu,\chi}$. We shall refer to these elements as ``pre-potentials''  since, in general, (like most ``potentials'' in physics) they define gauge equivalence classes of solutions to ``gauge-invariant'' equations. The ``gauge-invariant'' equations under consideration here include the source-free Maxwell system for the Minkowski-Maxwell field tensor, one of its modifications proposed by Bopp-Land\'{e}-Podolsky \cite{bopp,lande,podolsky}, the tensor equation for perturbations of Einstein's gravitational field equation in a matter-free background Minkowski spacetime and the electromagnetically-neutral massless Dirac equation. It will be shown how all these linear partial differential systems can be formulated in terms of certain differential tensor and spinor operators that render them amenable to analysis using the pre-potentials $\{\alpha,\Pi^{\nu,\chi}\}$. The class of particular solutions derived from these offers useful models of interest in laser physics and astrophysics. In particular, we indicate how to construct both Maxwell free-space models describing propagating finite-energy {\it multi-chiral} laser pulses that are bounded in all three spatial dimensions, and propagating gravitational pulses with similar characteristics. We argue that the former may offer new channels for quantum encryption and that the latter play a role in the formation of astrophysical jets observed in X-ray spectroscopy. The derivations of these models are discussed briefly since full details can be found in the indicated references. \\

Our formulation is given in terms of the language of differential geometry and section~\ref{sect:NOTATION} establishes essential definitions and our notational conventions. Section~\ref{sect:EM} describes the common technique used to solve the field systems above and draws upon ideas that seem to have originated in J. L. Synge's (1956) efforts to construct a classical model of the photon \cite{Synge}. Section~\ref{sect:EIN} sketches how a similar technique enables one to solve the linearised vacuum Einstein system for multi-chiral focussed gravitational pulses. In section~\ref{sect:SPINOR} we indicate how Minkowski spacetime spinor fields can be formulated in terms of sections of a $(3,1)$ Clifford algebra bundle and how the above pre-potential technique provides the necessary tools for deriving spinor solutions for the neutral, massless Dirac equation. Finally, in section~\ref{sect:CONC}, we summarise our conclusions.\\

\section{Definitions and Notational Conventions}\label{sect:NOTATION}
The natural mathematical language to discuss the differential properties of tensor fields on spacetime is in terms of differential forms and their associated exterior calculus \cite{benntucker}. In this section we give a brief summary of relevant notation used throughout the article.  \\

Let $\M$ denote spacetime modelled as a four-dimensional orientable Lorentzian manifold with {\it metric tensor field} $\g$ of signature $(3,1)\equiv (-,+,+,+)$. The metric tensor field can always be written in a local cobasis $\{e^{a}\}$ of differential $1$-forms as
\begin{align*}
	\g \,=\, \etaT_{ab}\,e^{a} \tensor e^{b} \,=\, -e^{0}\tensor e^{0} + e^{1}\tensor e^{1} + e^{2}\tensor e^{2} + e^{3}\tensor e^{3}
\end{align*}
with $\etaT_{ab}=\g(X_{a},X_{b})=\text{diag}(-1,1,1,1)$ in terms of the dual local basis $\{X_{a}\}$ of vector fields, defined so that $e^{a}(X_{b})=\delta^{a}_{b}$ ($a,b=0,1,2,3$). This induces the {\it inverse metric tensor field} 
\begin{align*}
	\g^{\inv} \,=\, \etaT^{ab}\,X_{a}\tensor X_{b} \,=\, -X_{0}\tensor X_{0} + X_{1}\tensor X_{1} + X_{2}\tensor X_{2} + X_{3}\tensor X_{3}
\end{align*}
where $\etaT^{ab}=\g^{\inv}(e^{a},e^{b})$ with $\etaT^{ab}\etaT_{bc}=\delta^{a}_{c}$. In our discussion of spinor fields below, based on a Clifford algebra, it will prove important to introduce the physically dimensionless tensor fields:
\begin{align}\label{scale_metric}
	\GG \,=\, \metscale\g \QQUAD{and} \GG^{\inv} \,=\, \frac{\g^{\inv}}{\metscale} 
\end{align}
where $\metscale$ is some chosen positive real number. \\

The tensor fields $\g$, $\g^{\inv}$ establish an isomorphism between the space of {\it vector fields} on $\M$ -- sections, denoted $\Gamma T\M$, of the {\it tangent bundle} $T\M$ -- and the space of {\it differential} $1${\it -forms} on $\M$ -- sections, denoted $\Gamma T^{*}\M$, of the {\it cotangent bundle} $T^{*}\M$. We represent this isomorphism with a `tilde':
\begin{alignat*}{2}
	\text{vector field}, X &\quad\longrightarrow\quad &\text{differential $1$-form}, &\,\wt{X}\,\equiv\, \g(X,-) \\[0.1cm]
	\text{differential $1$-form}, \alpha &\quad\longrightarrow\quad &\text{vector field}, &\,\wt{\alpha}\,\equiv\, \g^{\inv}(\alpha,-) 
\end{alignat*} 
where, for typographical economy, we use the same `tilde' symbol for both mappings. These are referred to as {\it metric duals} and the tilde map satisfies:
\begin{align*}
	\wt{\wt{X}} \,=\, X \QQUAD{and} \wt{\wt{\alpha}}\,=\, \alpha
\end{align*}  
for any vector field $X$ and differential $1$-form $\alpha$. Differential forms of arbitrary degree ($0\leq p \leq 4$) are identified with sections $\Gamma\Lambda\M$ of an {\it exterior bundle} $\Lambda\M$: a {\it differential }$p${\it -form} $\alpha$ is associated with a section of the $p$th exterior bundle $\Lambda^{p}\M$ whose elements are the totally antisymmetric sections of the type $(p,0)$ tensor bundle on $\M$ (hence $\alpha\in \Gamma\Lambda^{p}\M$). The bundle of differential $0$-forms is defined so that $\Gamma\Lambda^{0}\M\equiv\FM$, the smooth scalar functions on $\M$ and $\Gamma\Lambda^{1}\M\equiv\Gamma T^{*}\M$. Given a metric tensor field on $\M$, it is possible to relate the exterior algebra bundle $\Lambda\M$ with the bundle of Clifford algebras $C_{3,1}\M$. These algebras constitute the fibres of a {\it Clifford bundle} whose local sections define {\it Clifford elements} on $\M$. {\it Spinor fields} can be identified with local sections lying in certain minimal left ideals of this Clifford bundle. A Clifford calculus based on this Clifford bundle structure will be discussed in the last section on spinors. \\

Contraction of a differential $p$-form $\alpha$ with respect to a vector field $X$ is denoted $i_{X}\alpha$. The {\it interior operator} $i_{X}$ is a graded derivation defined by: 
\begin{align*}
	i_{X}(\alpha \w \beta) \,=\, i_{X}\alpha \w \beta + \eta\,\alpha \w i_{X}\beta \qquad\FORANY\quad \alpha\in\Gamma\Lambda^{p}\M, \beta\in\Gamma\Lambda^{q}\M
\end{align*}
in terms of the {\it involution}:
\begin{align*}
	\eta\,\alpha \,\equiv\, (-1)^{p}\,\alpha \qquad\FORANY\quad \alpha\in\Gamma\Lambda^{p}\M,
\end{align*}
and satisfying $i_{X}i_{X}\alpha=0$. If $p=1$ then one defines $i_{X}\alpha=\alpha(X)$ and if $p=0$: $i_{X}\alpha=0$. For an arbitrary $p$-form	$\alpha$, one has the useful identity:
\begin{align}\label{eaixalpha}
	p\,\alpha \,=\, e^{a} \w i_{X_{a}}\alpha
\end{align}
in terms of any arbitrary dual bases $\{e^{a}\}$ and $\{X_{a}\}$.\\

In the $\g$-orthonormal basis $\{e^{a}\}$ one has a canonical local $4$-form denoted:
\begin{align*}
	\mathcal{V} \,\equiv\, e^{0} \w e^{1} \w e^{2} \w e^{3}.
\end{align*}
The {\it Hodge map} $\star$ is a linear isomorphism mapping $p$-forms to $(4-p)$-forms on $\M$. It is defined by the relations:
\begin{align*}
	\alpha\in\Gamma\Lambda^{p}\M, X\in\Gamma T\M, f\in\FM:  
	\left\{	\begin{array}{rl}
				\star(\alpha \w \wt{X}) &\!=\; i_{X}\star\alpha \\[0.2cm]
				\star(\,f\alpha\, ) &\!=\; f\star\alpha \\[0.2cm]
				\star 1 &\!=\; \mathcal{V}.
			\end{array}.\right.
\end{align*}
In spacetime with signature $(3,1)$, the Hodge map and interior operator satisfy the following useful relations:
\begin{align}\label{HodgeProps}
	\alpha\in\Gamma\Lambda^{p}\M, X\in\Gamma T\M: 
	\left\{	\begin{array}{rl}
				\star\star\alpha 	&\!=\; -\eta\,\alpha \\[0.2cm]
				\star i_{X}\alpha &\!=\; -\star \alpha\w \wt{X}.
			\end{array}\right.
\end{align}
Since the Hodge map is an isomorphism it has an inverse denoted $\star^{-1}$:
\begin{align*}
	\star^{-1}\alpha \,=\,  -\eta\,\star \alpha \qquad \FORANY \quad \alpha\in\Gamma\Lambda^{p}\M.
\end{align*}
This implies $\star\star\alpha=-\eta\,\alpha$. The metric tensor field uniquely defines the {\it Levi-Civita covariant derivative operator} $\nabla_{X}$ with respect to any vector field $X$ satisfying:
\begin{alignat*}{2}
	\nabla_{X}(f_{1}Y+f_{2}Z) &\,=\!	\begin{array}[t]{rl}
											(\nabla_{X}f_{1})Y 	&\!\!\!+\; f_{1}\nabla_{X}Y \\[0.3cm]
																&\!\!\!+\; (\nabla_{X}f_{2})Z + f_{2}\nabla_{X}Z
										\end{array} &&\qquad  \raisebox{-0.3cm}{(\text{linearity})} \\[0.2cm]
	\nabla_{f_{1}X+f_{2}Y}Z &\,=\, f_{1}\nabla_{X}Z + f_{2}\nabla_{Y}Z &&\qquad (\text{$\mathcal{F}$-linearity}) \\[0.2cm]
	\nabla_{X}\left[\,\alpha(Y)\,\right] &\,=\, (\nabla_{X}\alpha)(Y) + \alpha(\nabla_{X}Y) &&\qquad (\text{commutes with contractions})\\[0.2cm]
	\nabla_{X}\g &\,=\, 0, &&\qquad (\text{metric compatibility}) \\[0.2cm]
	\nabla_{X}Y - \nabla_{Y}X &\,=\, [X,Y] &&\qquad (\text{torsion-free}) \\[0.2cm]
	\nabla_{X}\star &\,=\, \star\nabla_{X} &&\qquad (\text{commutes with the Hodge star})
\end{alignat*}
for any $X,Y,Z\in\Gamma T\M$,  $f_{1},f_{2}\in\FM$ and $\alpha\in\Gamma\Lambda^{1}\M$ where $[X,Y]\equiv XY-YX$ denotes the {\it Lie commutator bracket} (or simply {\it Lie bracket}) between $X$ and $Y$. In terms of the arbitrary local dual bases $\{e^{a}\}$ and $\{X_{a}\}$, any covariant derivative can be used to define a set of local {\it connection} $1$-{\it forms} $\{\om{a}{b}\}$ associated with each basis $\{X_{a}\}$ for $\Gamma T\M$:
\begin{align*}
	\nabla_{X_{a}}X_{b} \,=\, \om{c}{b}(X_{a})\,X_{c}.
\end{align*}
Since $\nabla_{X_{a}}$ commutes with contractions:
\begin{align}\label{DELea}
	\nabla_{X_{a}}e^{c} \,=\, -\om{c}{b}(X_{a})\,e^{b}.
\end{align}
The {\it Christoffel symbols} $\{\Gamma_{ab}{}^{c}\}$ are components of $\nabla_{X_{a}}X_{b}$ in this basis: 
\begin{align*}
	\nabla_{X_{a}}X_{b} \,=\, \Gamma_{ab}{}^{c}\,X_{c}
\end{align*}
and hence 
\begin{align*}
	\om{a}{b} \,=\, \Gamma_{cb}{}^{a}e^{c}.
\end{align*}
Since a Levi-Civita covariant derivative is torsion-free one has, in any $\g${\it -orthonormal} basis
\begin{align*}
	\omega_{ab} \,=\, -\omega_{ba}
\end{align*}
where $\omega_{ab}=\g_{ac}\,\om{c}{b}$. We define the {\it Levi-Civita covariant differential operator} on tensor fields:
\begin{align}\label{TJW_LEVI_DIFF_OP}
	\nabla \,\equiv\, e^{a} \tensor \nabla_{X_{a}}.
\end{align}
The covariant derivative operator associated with $\g$ is used to define the {\it curvature operator of} $\nabla$:
\begin{align*}
	\pmb{R}_{X,Y}\,\equiv\, [\nabla_{X},\nabla_{Y}] - \nabla_{[X,Y]} \qquad \FORALL\quad X,Y\in\Gamma T\M.
\end{align*}
This operator defines the $(3,1)$ {\it curvature tensor field} $\RIEM$ of $\nabla$:
\begin{align*}
	\RIEM(X,Y,Z,\beta) \,=\, \beta(\pmb{R}_{X,Y}Z), \qquad \FORALL\quad X,Y,Z\in\Gamma T\M, \beta\in\Gamma\Lambda^{1}\M.
\end{align*}	
Contracting the curvature tensor field yields the $(2,0)$ {\it Ricci tensor field} $\texttt{Ric}$ of $\nabla$:
\begin{align*}
	\texttt{Ric}(X,Y) \,=\, \RIEM(X_{a},X,Y,e^{a}), \qquad\FORALL\quad X,Y\in\Gamma T\M
\end{align*}
where the arbitrary bases $\{X_{a}\}$, $\{e^{a}\}$ are dual. The Ricci tensor can be {\it contracted} to construct the {\it Ricci curvature scalar} $\mathcal{R}$:
\begin{align*}
	\mathcal{R} \,\equiv\,\texttt{Ric}(X_{a},X^{a})
\end{align*}
where $X^{a}=\g^{ab}X_{b}$.\\

While $\nabla_{X}$ has a type-preserving action on any tensor field, the {\it exterior derivative} $d$ is defined only to act on antisymmetric tensor fields (differential forms) and is nilpotent: $d\circ d=0$. Provided $p+q<4$, the exterior derivative satisfies
\begin{align*}
	d( \alpha \w \beta) \,=\, d\alpha \w \beta + \eta\,\alpha \w d\beta, \qquad \FORANY\quad \alpha\in\Gamma\Lambda^{p}\M,  \beta\in\Gamma\Lambda^{q}\M,
\end{align*}
otherwise $d(\alpha \w \beta)=0$. Given an arbitrary local cobasis $\{e^{a}\}$ with dual basis $\{X_{a}\}$ one may verify the useful identity:
\begin{align}\label{dDEL}
	d \,=\, e^{a} \w \nabla_{X_{a}}.
\end{align}
From these relations and the nilpotency of $d$, it follows immediately that the {\it co-derivative} $\delta$ defined by
\begin{align}\label{defCODERIV}
	\delta \alpha \,\equiv\, \star^{-1}d\star \eta\,\alpha  \qquad\FORANY\quad \alpha\in\Gamma\Lambda^{p}\M.
\end{align}
is also nilpotent: $\delta\circ\delta =0$. Unlike $d$, however, $\delta$ is not a graded derivation on differential forms. On spacetime one has $\delta\alpha = \star d \star \alpha$ for any $p$-form $\alpha$. Using (\ref{dDEL}) and the properties above, one may verify the relation:
\begin{align}\label{deltaDEL}
	\delta \,=\, -i_{X^{a}}\nabla_{X_{a}}
\end{align}
where $X^{a}=\wt{e^{a}}=\g^{\inv}(e^{a},-)=\etaT^{ab}X_{b}$. \\

The Lie bracket gives rise to an $\RR${\it -linear derivation} $\Lie_{X}$ on functions $f,g\in\FM$:
\begin{align*}
	[X,Y] \,\equiv\,\Lie_{X}Y, \qquad \Lie_{X}f \,=\, Xf, \qquad \Lie_{cX}Y\,=\, c\Lie_{X}Y, \qquad \Lie_{X}(fg) \,=\, (\Lie_{X}f)g + f(\Lie_{X}g)
\end{align*}
for $c\in\RR$. The operator $\Lie_{X}$ is called the {\it Lie derivative with respect to $X$} and can be generalised to act on arbitrary tensor fields:
\begin{align}\label{LieST}
	\Lie_{X}(\texttt{S} \tensor \texttt{T}) \,=\, \Lie_{X}\texttt{S} \tensor \texttt{T} + \texttt{S} \tensor \Lie_{X}\texttt{T}
\end{align}
for any $X\in\Gamma T\M$ and arbitrary tensor fields $\texttt{S},\texttt{T}$. In particular, it commutes with contractions and satisfies:
\begin{align*}
	\Lie_{X}(f\texttt{T}) &\,=\, (Xf)\texttt{T} + f\Lie_{X}\texttt{T}.
\end{align*}	
From (\ref{LieST}), it is also a derivation on the exterior algebra of differential forms:
\begin{align*}
	\Lie_{X}(\alpha \w \beta) \,=\, \Lie_{X}\alpha \w \beta + \alpha \w \Lie_{X}\beta
\end{align*}
for arbitrary $X\in\Gamma T\M$, $\alpha\in\Gamma\Lambda^{p}\M$ and  $\beta\in\Gamma\Lambda^{q}\M$. From the definitions above, one has the identities \cite{benntucker}:
\begin{alignat}{2}
	\label{dLIE}		d\Lie_{X} &\,=\, \Lie_{X}d \qquad &&(\text{commutes with exterior derivative})\\[0.2cm]
	\nonumber			\Lie_{X}\left[\,\alpha(Y)\,\right] &\,=\, (\Lie_{X}\alpha)(Y) + \alpha(\Lie_{X}Y) \qquad &&(\text{commutes with contraction}) \\[0.2cm]
	\label{LieCartan}	\Lie_{X} &\,=\, i_{X}d + di_{X} \qquad &&(\text{Cartan's magic formula})
\end{alignat}
for any $X,Y\in\Gamma T\M$, arbitrary tensor fields $\texttt{T},\texttt{S}$, $f\in\FM$ and $\alpha\in\Gamma\Lambda^{1}\M$. \\

A spacetime $\M$ is said to admit a {\it conformal Killing vector field} $\mathcal{K}$ if
\begin{align*}
	\Lie_{\mathcal{K}}\g \,=\, 2\nu\,\g, \qquad \nu\in\FM, \,\nu> 0. 
\end{align*}
If $\nu=0$, the vector field $\mathcal{K}$ is a {\it Killing vector field} $K$ with
\begin{align}\label{KILLING}
	\Lie_{K}\g \,=\, 0. 
\end{align}
It follows from (\ref{KILLING}) that the Lie derivative with respect to any Killing vector field $K$ commutes with the Hodge star map:
\begin{align}\label{LieKstar}
	\Lie_{K}\star \,=\, \star\Lie_{K}.
\end{align}
Furthermore, if $\nu\neq 0$ for any metric tensor field on a manifold with dimension $2m$, one has for any $m$-form $F$ the identity:
\begin{align*}
	\Lie_{\mathcal{K}}\left(\star F\right) \,=\, \star \left(\,\Lie_{\mathcal{K}}F\,\right).
\end{align*}
This relation has relevance to the conformal symmetry of source-free Maxwell equations for the Maxwell $2$-form $F$ in spacetime where $m=2$. \\

\section{Electromagnetic Pulses in Spacetime Vacua}\label{sect:EM}
The method of pre-potentials can be used to construct solutions to the source-free linear vacuum Maxwell system in Minkowski spacetime:
\begin{align*}
	dF \,=\, 0 \QQUAD{and}	\delta F \,=\, 0
\end{align*}
where $F$ denotes the electromagnetic $2$-form. The first equation is satisfied locally on some open spacetime domain by $F=dA$ for some {\it electromagnetic potential} $1$-form $A$. The second equation then becomes:
\begin{align}\label{SFM2}
	\delta dA \,=\, \star d \star dA \,=\, 0
\end{align}
using (\ref{defCODERIV}). We seek particular local {\it complex solutions} to (\ref{SFM2}) with the ans\"{a}tz:
\begin{align}\label{ALP}
	A \,=\, \star\,d\left(\alpha\Pi\right)
\end{align}
for the electromagnetic potential in terms of a {\it complex scalar pre-potential} $\alpha$ and a {\it complex} $2$-form\footnote{When the $2$-form $\alpha\Pi$ is {\it real} it reduces to the Hodge dual of a {\it real Hertz potential} for a Maxwell system.} $\Pi$. The $2$-form $\Pi$ is chosen so that, in general, it is both locally {\it closed} and {\it co-closed}:
\begin{align}\label{PiPROPS}
	d\Pi \,=\, 0 \QQUAD{and} \delta\Pi \,=\, 0
\end{align}
respectively. It follows from (\ref{dDEL}) and (\ref{deltaDEL}) that if $\nabla_{X_{a}}\Pi=0$ for all basis elements in $\{X_{a}\}$ the form $\Pi$ is both closed and co-closed (such a $\Pi$ is said to be locally {\it parallel} or {\it covariantly constant}). Using (\ref{HodgeProps}), (\ref{dDEL}), (\ref{deltaDEL}) and (\ref{PiPROPS}), the ans\"{a}tz (\ref{ALP}) yields:
\begin{align}\label{stardA}
	\star dA \,=\, \delta d\alpha \w \Pi  + \nabla_{X_{a}}d\alpha \w i_{X^{a}}\Pi.
\end{align}
Using (\ref{eaixalpha}) and (\ref{DELea}), one has the identity:
\begin{align*}
	\nabla_{X_{a}}\beta \,=\, i_{X_{a}}d\beta + di_{X_{a}}\beta - \om{c}{a} \w i_{c}\beta \qquad\FORANY\quad \beta\in\Gamma \Lambda^{p}\M.
\end{align*}
This can be applied to simplify the last term in (\ref{stardA}) yielding:
\begin{align*}
	\star dA \,=\, \delta d\alpha \w \Pi  + d\left(i_{X_{a}}d\alpha \w i_{X^{a}}\Pi\right)
\end{align*}
where (\ref{PiPROPS}) is also used. From the nilpotency of $d$ and (\ref{PiPROPS}) one has:
\begin{align*}
	d\star F \,=\, d \star d A \,=\, d \delta d \alpha \w \Pi .
\end{align*}
Hence a particular sub-class of solutions to (\ref{SFM2}) can be found when $\alpha$ satisfies the spacetime scalar wave equation:
\begin{align}\label{BOX}
	\Box\,\alpha \,\equiv\, \delta d \alpha \,=\, 0.
\end{align}
Thus, {\it any} solution to (\ref{BOX}) for the pre-potential $\alpha$ can, given a covariantly constant $2$-form $\Pi$, be used to construct a solution to the source-free Maxwell system using (\ref{ALP}) and $F=dA$. Of course, (\ref{BOX}) has many solutions, however with the advent of the laser {\it non-singular spatially bounded solutions in all three dimensions} are of particular interest as models for the electromagnetic fields in modern single-cycle laser devices.  \\

In a Cartesian chart with local {\it dimensionless} coordinates $\{x^{\mu}\}\equiv\{t,x,y,z\}$ the {\it dimensionless} Minkowski spacetime metric tensor field $\g$ is given by 
\begin{align*}
	\left.\g\right|_{\text{Cart}} \,=\, - dt \tensor dt + dx \tensor dx  + dy \tensor dy + dz \tensor dz.
\end{align*}
A particularly simple class of such solutions can be generated from the complex axially-symmetric scalar field \cite{Synge,Britt,Visser,Ziol85,Ziol89}
\begin{align}\label{alp}
   \alpha(t,x,y,z) \,=\, \frac{\W^{2}}{x^{2} + y^{2} + [\,a + i(z-t)\,]\,[\,b-i( z+t )\,]  }
\end{align}
where $\W,a,b$ are real constants. When $a,b$ are {\it strictly positive} constants, (\ref{alp}) defines a non-singular solution to $\Box\,\alpha=0$. This scalar field is a non-separable solution to (\ref{BOX}) with the relative sizes of $a$ and $b$ determining both the direction of propagation along the $z$-axis of the dominant peak of the real or imaginary part of the scalar field and the number of oscillations in its peak. When $a \gg b$, the dominant peak propagates along the $z$-axis to the right. The parameter $\W$ determines the magnitude of such a peak. \\

To construct a source-free vacuum Minkowski Maxwell solution, we must also define a {\it covariantly constant} $2$-form. One may define a {\it complex six-dimensional chiral eigen-basis of covariantly constant $2$-forms} $\Pi^{s,\,\chi}$ satisfying 
\begin{align}\label{RWT_C_13}
	\frac{1}{i}\,\Lie_{\partial/\partial\phi}  \, \Pi^{s,\,\chi} \,=\, \chi\,\Pi^{s,\,\chi }
\end{align}
with $s \in \{\mathrm{CE}, \mathrm{CM}\}$, $\chi \in \{ {1}, 0, -1\}$ where $x=\rho\cos(\phi)$, $y=\rho\sin(\phi)$. Such a basis takes the form:
\begin{align}\label{PIEQNS}
	\left\{	\begin{array}{rl}
				\Pi^{\mathrm{CE},\pm1}		&\!=\;	d(x\pm i y) \w d t \,=\, e^{\pm i\phi}\left( d\rho \pm i\rho\,d\phi \right) \w dt ,\\[0.2cm]
				\Pi^{\mathrm{CE},0}			&\!=\;	d z \w d t,\\[0.2cm]
				\Pi^{\mathrm{CM},\chi }		&\!=\; \star\, \Pi^{\mathrm{CE},\chi }
			\end{array}\right.
\end{align}
The index $s$ indicates that the CE (CM) chiral family contain electric (magnetic) fields that are orthogonal to the $z$-axis when $\chi = 0$. Furthermore, the rationale for the $\chi$-labelling of the $2$-form solution
\begin{align}\label{Fschi}
	F^{s,\chi} \,=\, d\left[\,\star\,d(\alpha\Pi^{ s, \chi} )\,\right]
\end{align}
in this chiral basis follows from an exploration of the properties of spacetime dimensionless proper-time $\tau$ parameterised curves $x^{\mu}=C^{\mu}(\tau)$ of massive test particles interacting with such electromagnetic fields according to the covariant Lorentz force equations of motion:
\begin{align}\label{EOM}
	\left\{	\begin{array}{rl}
				\ds\left.\nabla_{\dot{C}(\tau)}\dot{C}(\tau)\right|_{C(\tau)} &\!=\; \ds\left.\frac{q}{m_{0}}\, \wt{i_{\dot{C}(\tau)}F}\,\right|_{C(\tau)} \\[0.6cm]
				\ds\left.\g(\dot{C}(\tau),\dot{C}(\tau))\right|_{C(\tau)} &\!=\; -1
			\end{array}\right.
\end{align}
where $\dot{C}(\tau)\equiv C_{\star}(\partial/\partial\tau)|_{C(\tau)}$ denotes the tangent vector of $C(\tau)$ and $q,m_{0}$ are dimensionless charge and mass parameters respectively. \\

\LASERSPACECURVES{%
	Three-dimensional space-curves for particles subject to an incident $(\text{CM},1)$ laser pulse (left), $(\text{CM},0)$ laser pulse (centre) and $(\text{CM},-1)$ laser pulse (right). Each particle has initial velocity in the $z$-direction. The shaded circular disc region indicates the initial laser spot size relative to the black markers on the space-curves that denote the initial positions of the charged test particles.
	}
	
In terms of dimensionless Minkowski cylindrical polar coordinates $\{ t,\rho,\phi,z \}$ the metric tensor field $\g$ takes the form:
\begin{align*}
	\left.\g\right|_{\text{polar}} \,=\, -dt \tensor dt + d\rho \tensor d\rho  + \rho^{2}\,d\phi \tensor d\phi + dz \tensor dz.
\end{align*}
A choice of values for the dimensionless parameters $\{\W,a,b,q,m_{0}\}$ can be used to solve (\ref{EOM})   numerically  for a collection of trajectories for charged particles, each arranged initially around the circumference of a circle in a plane orthogonal to the propagation axis of incident CM type electromagnetic pulses with different chirality. The resulting space-curves in $3$-dimensions, displayed in figure~\ref{fig:laserspacecurves} clearly exhibit the different responses of charged matter to CM pulses with distinct chirality values \cite{GTW_Lasers}.  Similar space-curves arise from charged particles interacting with chiral CE type modes.\\

The {\it dimensionless} energy, linear and angular momentum parameters for the pulse in vacuo can be calculated from {\it the Maxwell drive} $3${\it -forms} $\tau_{K}$ \cite{benntucker,TW_Magneto,TW_Proc}:
\begin{align*}
	\tau_{K} \,=\, \frac{1}{2}\left(i_{K}\realF \w \star \realF - \realF \w i_{K}\star \realF\right)
\end{align*}
where $\realF=\text{Re}(F)$ and $K$ represents a Killing vector field generating time, space and axially symmetric rotations respectively.  With $K=\partial/\partial t$ a timelike Killing vector field on Minkowski spacetime, the total {\it physical} pulse electromagnetic energy associated with a spatially bounded pulse in all three dimensions can be calculated using:
\begin{align*}
	\mathcal{E}[\mathcal{S}] \,=\, E_{0}\int_{-\infty}^{\infty}dt\int_{\mathcal{S}}\,\mathcal{J}_{K}(t,\rho,\phi,z) 
\end{align*}
where $E_{0}$ is a real positive constant with the physical dimensions of energy, $\mathcal{S}$ is any plane with constant $z=z_{0}\geq 0$ and $\mathcal{J}_{K}\equiv -i_{K}\tau_{K}$ is the dimensionless {\it energy current} $2$-{\it form} \cite{TW_Magneto}, related to the {\it Poynting vector field}. We can then write
\begin{align*}
	\mathcal{E}[\mathcal{S}] \,=\, E_{0}\int_{-\infty}^{\infty}dt\,\int_{0}^{\infty}d\rho\,\int_{0}^{2\pi}d\phi\;P(t,\rho,\phi)
\end{align*}
in terms of a {\it  total power density profile} $0$-form $P$ defined by
\begin{align*}
	P(t,\rho,\phi) \,\equiv\, \left.i_{\partial/\partial\phi}\,i_{\partial/\partial \rho}\,\mathcal{J}_{K}(t,\rho,\phi,z)\right|_{z=z_{0}}.
\end{align*}
Figure~\ref{fig:lasercompact} displays a clearly pronounced principle maximum in $P$ as a function of $x=\rho\cos(\phi)$, $y=\rho\sin(\phi)$ at $z=0$, $t=0$ for the specific choice of parameters used to produce the three-dimensional space-curves in figure~\ref{fig:laserspacecurves}. \\

\LASERCOMPACT{%
	Dimensionless power profile for a (CM,$1$) laser pulse indicating the spatial boundedness of the pulse in space.  
	}
	
A similar analysis using spacelike translational or rotational Killing vectors can be used to determine the total integrated flux of linear or angular momentum current (total Newtonian force or torque) respectively. We deduce that the pulse momentum and angular momentum in the propagation direction can transfer an impulsive force and torque respectively to charges lying in an orthogonal plane. More generally, the classical configurations of a high energy pulse labelled CE or CM could be distinguished experimentally by its interaction with different arrangements of charged matter. Particular choices of pulse parameters $\{\W,a,b\}$ can be made to model the physical characteristics of existing laser pulses and have been used, for electromagnetic micropulses of duration $t_{0}$ with $\mathcal{E}t_{0}\lesssim \hbar$, to model an {\it effective} description of general non-stationary quantum states of a laser pulse \cite{GTW_Lasers}. It is suggested that such a description may have utility for simulating a novel transfer of quantum information and may be worthy of further  investigation for technological applications such as quantum computing and encryption.\\

Minkowski spacetime admits Killing vector fields that generate the group of spatial and temporal translations, as well as spatial rotations and Lorentzian boosts (hyperbolic rotations). For example, in the Cartesian chart the Killing vector field $K_{z}=\partial/\partial z$ generates unit translations in the direction $\partial/\partial z$ while the Killing vector field:
\begin{align}\label{DEF_Kphi}
	K_{\phi} \,=\, x\,\frac{\partial}{\partial y} - y\,\frac{\partial}{\partial x} \,=\, \frac{\partial}{\partial \phi}
\end{align}	
generates $O(2)$ rotations about that direction and the Killing vector field $K_{t}=\partial/\partial t$ generates unit timelike translations. \\

From the definition (\ref{HodgeProps}), if $K$ is any local Killing vector field it commutes with the Hodge map and the exterior derivative $d$ on arbitrary differential forms. Hence it also commutes with the coderivative $\delta$ on forms.  So, for any Minkowski spacetime Maxwell solution $F$ with source $1$-form $j$ satisfying $\delta F = j$ and $dF=0$, one has:
\begin{align*}
	\delta\left( \Lie_{K}F \right) \,=\, \Lie_{K}j \QQUAD{and} d\left(\Lie_{K}F\right) \,=\, 0
\end{align*}
i.e. $\Lie_{K}F$ is also a Maxwell solution with source $1$-form $\Lie_{K}j$. In any {\it source-free domain} $\UU\in\M$, this implies that in a Cartesian chart one can generate new source-free Maxwell solutions in $\UU$:
\begin{align*}
	F^{(p,q,r)} \,=\, \Lie_{\partial/\partial x}^{p}\,\Lie_{\partial/\partial y}^{q}\,\Lie_{\partial/\partial z}^{r}\,F, \qquad p,q,r \,=\, 0,1,2,\ldots
\end{align*}
in terms of any $2$-form $F$ satisfying $dF=0$, $\delta F=0$ in $\UU$ and where
\begin{align*}
	\Lie_{K}^{p}F \,\equiv\, \underbrace{\Lie_{K}\Lie_{K} \cdots \Lie_{K}}_{\text{$p$ times}}\,F.
\end{align*}
Such solutions are proportional to electrostatic Cartesian electric point multipoles if $F$ is static ($\Lie_{\partial/\partial t}F=0$) and describes the electric monopole (Coulomb) solution at some point located in some region of spacetime not belonging to $\UU$. Similarly, it follows from (\ref{dLIE}), (\ref{LieCartan}) and (\ref{LieKstar}) that the operator $\Lie^{r}_{K_{\pm}}$, where 
\begin{align}\label{Kpm}
	K_{\pm} \,=\, \g^{\inv}\!\left(d(\,\rho e^{\pm i\phi}),-\right) \,=\, e^{\pm i\phi}\left(\frac{\partial}{\partial \rho} \pm \frac{i}{\rho}\frac{\partial}{\partial \phi}\right)
\end{align}
is a complex Killing vector ($\Lie_{K_{\pm}}\g=0$) generating translations in the transverse $(x,y)$-plane, defines the $2$-form 
\begin{align*}
	F^{s,\chi\pm r} \,\equiv\, \Lie^{r}_{K_{\pm}}F^{s,\chi} \QQUAD{satisfying} \frac{1}{i}\Lie_{\partial/\partial\phi}F^{s,\chi\pm r} \,=\, (\chi\pm r)\,F^{s,\chi\pm r} \qquad(r=0,1,2,\ldots)
\end{align*}
describing a family of {\it multi-chiral} source-free Maxwell solutions including those that exhibit a non-singular, propagating, non-stationary pulse-like nature, spatially bounded in all three dimensions. \\

As a further application of pre-potentials to the solution of a linear system of differential equations, we briefly shift attention to generalizations of Maxwell's theory. Motivated by the desire to ameliorate the divergences in perturbative QED a number of generalized theories of electromagnetism have been proposed. To date, there is little experimental evidence for testing their predicted departures from Maxwell's theory. However, with the increase in laser technology one may now be entering regimes that can discriminate between such theories. In particular, {\it Bopp-Land\'{e}-Podolsky electrodynamics} are {\it linear} in the electromagnetic fields, contain a fundamental real non-zero constant $\lambda_{0}$ and approach Maxwell's theory when this constant tends to zero. Unlike Maxwell's theory, they contain solutions describing static point charges with \textit{finite} electromagnetic energy. \\

The classical source-free Bopp-Land\'{e}-Podolsky field equation in vacuo for a complex electromagnetic field tensor $F=dA$ is \cite{Tucker_BP}:
\begin{align}\label{BLP}
	d\star F - \lambda_{0}^{2}\,d\delta d\star F \,=\, 0.
\end{align}
Clearly, all classical source-free vacuum Maxwell solutions satisfy this equation. However, there exist additional solutions to (\ref{BLP}) that are \textit{not} source-free vacuum Maxwell solutions. For example, such additional solutions of (\ref{BLP}) follow from (\ref{ALP}) provided:
\begin{align}\label{BLPalpha}
	\Box\,\alpha + \lambda_{0}^{2}\,\Box^{2}\alpha \,=\, 0
\end{align}
when supplemented by a closed and co-closed choice of $\Pi$. Furthermore, it follows that as $\lambda_{0}\rightarrow 0$, one recovers the class of Maxwell solutions discussed above. For non-zero $\lambda_{0}$, there exist particular non-Maxwellian solutions to (\ref{BLPalpha}) satisfying
\begin{align*}
	\Box\,\alpha + \cfrac{1}{\lambda_{0}^{2}}\, \alpha \,=\, 0, \qquad \lambda_{0}\,\neq\, 0.
\end{align*}
Thus {\it any} solution to this equation for the pre-potential $\alpha$ can, given {\it any} covariantly constant $2$-form $\Pi$, be used to construct a solution to the source-free  Bopp-Land\'{e}-Podolsky system using (\ref{ALP}) and $F=dA$. \\

In dimensionless Minkowski cylindrical coordinates $\{t,\rho,\phi,z\}$, a particular non-separable axially symmetric spatially bounded solution in three dimensions is given for arbitrary non-zero real dimensionless $\Lambda_{0}$ by:
\begin{align*}
	\alpha(t,\rho,z) \,=\, \frac{\Lambda_{0}^{2}}{ \zeta(t,\rho,z) }\,K_{1}\!\left( \frac{\zeta(\rho,t,z)}{\lambda_{0}} \right)
\end{align*}
where 
\begin{align*}
	\zeta(t,\rho,z) \,\equiv\, \sqrt{ \rho^{2} + \left[a + i(z-t)\right]\left[b - i(z+t)\right]}
\end{align*}
in terms of strictly positive (real) parameters $a$, $b$ that determine pulse propagation along the positive $z$-axis and where $K_{1}(z)$ denotes the first order modified Bessel function of the second kind. The function $\zeta(t,\rho,z)$ is non-zero for all real $t,\rho,z$ and hence both the real and imaginary parts of $\alpha(t,\rho,z)$ are bounded. To our knowledge, this constitutes a new source-free finite-energy solution for $F$ to Bopp-Land\'{e}-Podolsky electrodynamics that does not arise in classical Maxwell electrodynamics \cite{GTW_PIPAMON}. \\

The interaction of the Bopp-Land\'{e}-Podolsky pulse with classical charged test particles follows from the divergence of the Bopp-Land\'{e}-Podolsky stress-energy-momentum tensor \cite{Tucker_BP}, and may offer an experimental means to discriminate between the Maxwell and Bopp-Land\'{e}-Podolsky descriptions of source-free electromagnetic fields in vacuo. \\

\section{Gravitational Pulses in Vacua}\label{sect:EIN}
The pre-potential approach can also be used to analyse gravitational field equations. In particular, it can be used to generate solutions to the {\it linearised Einstein vacuum system}. In any matter-free domain of spacetime $\UU\subset\M$, an Einsteinian gravitational field is described by a real symmetric covariant rank-two metric tensor field $\wh{\g}$ with Lorentzian signature that satisfies the vacuum Einstein equation 
\begin{align}\label{EINgEQS}
	\EIN{\wh{\g}} \,=\, 0 
\end{align}	
where
\begin{align}\label{VAC}
	\EIN{\wh{\g}} \,\equiv\, \RIC{\wh{\g}} - \frac{1}{2}\,\TR{\wh{\g}}(\RIC{\wh{\g}})\,\wh{\g}
\end{align}
and $\RIC{\wh{\g}}$ is the Levi-Civita Ricci tensor associated with the torsion-free, metric-compatible Levi-Civita connection $\NABLA{\wh{\g}}$. By taking the $\wh{\g}$-trace of (\ref{EINgEQS}), one also has $\RIC{\wh{\g}}=0$. A coordinate independent linearisation of (\ref{VAC}) about an arbitrary Lorentzian metric  can be found in \cite{stewart,TuckerGEM}. In particular, a linearisation about a real flat Minkowski spacetime metric $\etaT$ on $\mathcal{U}$ determines the linearised metric
\begin{align*}
	\g \,=\, \etaT + \h
\end{align*} 
and to first order one writes $\wh{\g} = \g + O(\kappa^{2}).$  The real variable $\kappa$ is a dimensionless parameter in $\h$ used to keep track of the expansion order and
\begin{align}\label{ETA}
	\etaT \,=\, -\, e^{0} \tensor e^{0} + \sum_{k=1}^{3} e^{k} \tensor e^{k} \;=\; \etaT_{ab}\,e^{a}\tensor e^{b}
\end{align}
in {\it any} real $\etaT$-orthonormal coframe\footnote{%
	An arbitrary coframe on $\mathcal{U}$ is a set of $1$-forms  $\{e^{a}\}$ satisfying $e^{0}\w e^{1}\w e^{2}\w e^{3} \neq 0 $.  If $\beta=\beta_{ab}e^{a}\tensor e^{b}$ in {\it any} coframe $\{e^{a}\}$ on $\mathcal{U}$, $\TR{\eta}(\beta)\equiv\beta_{ab}\eta^{ab}$ with $\eta^{ab}\eta_{bc}=\delta^{a}_{c}$.%
	} 
on $\UU$. Since we explore the source-free Einstein equation (relevant to the motion of test-matter far from any sources) the scale associated with any linearised solutions must be fixed by the solutions themselves rather than any coupling to self-gravitating matter. Furthermore since only dimensionless relative scales have any significance we define the real tensor field $\h$ to be a {\it perturbation} of $\etaT$ on $\UU$ relative to {\it any} local $\etaT$-orthonormal coframe $\{e^{a}\}$  provided 
\begin{align}\label{CNDS}
	|\,\h(X_{a},X_{b})\,| \,<\, 1 \QQUAD{on \;$\mathcal{U}$} \text{for all $a,b=0,1,2,3$}
\end{align} 
where  $e^{a}\in \Gamma T^{*}\mathcal{U}$, $X_{b}\in \Gamma T\mathcal{U}$, $e^{a}(X_{b})=\delta^{a}_{b}$. It should be noted that the perturbation order of any component of the $\etaT$-covariant {\it derivative} of a tensor and its $\etaT$-trace relative to such a coframe is not necessarily of the same order as that assigned to the tensor. Thus perturbation order is not synonymous with ``scale'' in this context. We use the conditions (\ref{CNDS}) to {\it define} perturbative Lorentzian spacetime to be sub-domains $\UUP\subset \UU$ where 
\begin{align}\label{perturb_domains}
	\mathop{\text{max}}_{a,b} \,\vert \h(X_{a},X_{b})\vert \,<\, 1.
\end{align}
The tensor $\h\equiv\TRREV{\etaTs}(\psiT')$ with $\psiT'\equiv\mathrm{Re}(\psiT)$ may be constructed \cite{TuckerGEM} from any complex covariant symmetric rank two tensor $\psiT$ satisfying the {\it linearised source-free Einstein equation on the perturbative domain} $\UUP$:
\begin{align}\label{LIN_EIN_SF}
	\LAP{\etaTs}(\psiT) - 2\TRREV{\etaTs}\left(\text{Sym}\NABLA{\etaTs}(\,\DIV{\eta}(\psiT)\,) \right) \,=\, 0.
\end{align}
Here and below, $\NABLA{\etaTs}$ denotes the operator of Levi-Civita covariant differentiation associated with $\etaT$, $X^{a}\equiv\etaT^{ab}X_{b}$, $Y\equiv\NABLA{\etaTs}_{X_{a}}X^{a}$, for any tensor $\T$ on $\UU$:
\begin{align}
	\label{TJW_LAP_T} \LAP{\etaTs}(\T)	&\,\equiv\, \NABLA{\etaTs}_{Y}\T - \NABLA{\etaTs}_{X_{a}}\NABLA{\etaTs}_{X^{a}}\T	\\[0.2cm]
	\nonumber \TRREV{\etaTs}(\T)	&\,\equiv\, \T - \frac{1}{2}\TR{\etaTs}(\T)\,\etaT 
\end{align}
and for all complex covariant symmetric rank-two tensors $\T$ on $\UU$: 
\begin{align}\label{TJW_DEF_DIV}
	\DIV{\etaTs}(\T)	&\,\equiv\, (\NABLA{\etaTs}_{X_{a}}\T)(X^{a},-) .
\end{align}
It follows from (\ref{TJW_LAP_T}) that for any scalar field $\alpha$:
\begin{align}\label{TJW_LAP_ALP}
	\LAP{\etaTs}(\alpha) \,=\, -\delta d\alpha.
\end{align}
Since for any $\g$ the trace-reverse map $\TRREV{\g}$ satisfies $\TRREV{\g}\circ\TRREV{\g}=\text{Id}$, if $\psiT'$ is trace-free with respect to $\etaT$, then $\h=\psiT'$. If $\psiT$ is also divergence-free with respect to $\etaT$, then $\LAP{\etaTs}(\psiT)=0$. Thus, divergence-free, trace-free solutions $\psiT$ satisfy:
\begin{align}\label{DFTF_LAP_EQ}
	\LAP{\etaTs}(\psiT) \,=\, 0 \QQUAD{and}	\DIV{\etaTs}(\psiT) \,=\, 0.
\end{align}
We now construct complex solutions $\psiT$ to (\ref{DFTF_LAP_EQ}) in terms of a complex $0$-form pre-potential $\alpha$. These can then be used to generate a real linearised metric $\g$. The properties of the scalar field $\alpha$ needed to generate chiral, wave-like and pulse-like solutions to the linearised source-free Einstein equations have been developed in \cite{TW_GPulses}. To emulate the methodology used above for the electromagnetic pulse solutions we note that a key role is played by the commutativity of certain exterior operators with covariantly constant {\it antisymmetric} tensors and the Hodge de-Rham Laplacian $\delta d$ on scalar fields in Minkowski spacetime. In Einsteinian gravitation one seeks similar properties involving the tensor Laplacian operator $\LAP{\g}$, $\nabla^{(\g)}$ and {\it symmetric} tensors in spacetimes with a metric tensor $\g$. The key general identity is the relation between $\LAP{\g}$, $\nabla^{(\g)} $ and the curvature operator $\pmb{R}^{(\g)}_{X,Y}$ of the torsion-free, metric compatible connection $\nabla^{(\g)} $:
\begin{align*}
	\LAP{\g} \left(\nabla^{(\g)} \alpha\right) \,=\,\nabla^{(\g)}\LAP{\g}(\alpha) + \pmb{R}^{(\g)}_{X_{j},\wt{d \alpha}} \,e^{j}
\end{align*}
valid for any scalar field $\alpha$. On flat Minkowski spacetime with metric $\etaT$, this yields the commutation relation\footnote{%
	For any scalar field $\alpha$  and metric tensor $\g$, $\NABLA{\g}\alpha=d\alpha$ and is independent of $\g$.%
	} :
\begin{align}\label{LAP_COMM_REL}
	[\LAP{\etaTs},\nabla^{(\etaTs)}]\alpha \,=\, [\LAP{\etaTs},d]\alpha \,=\, 0 \qquad\text{on scalar fields $\alpha$}.
\end{align}
Thus, if $\alpha$ is {\it any}  complex (four times differentiable)  scalar field on $\mathcal{U}$ then 
\begin{align}\label{TJW_LAP}
	\NABLA{\etaTs} \NABLA{\etaTs}\,\LAP{\etaTs}(\alpha) \,=\, \LAP{\etaTs}\!\left(\NABLA{\etaTs} \NABLA{\etaTs}\alpha\right) \,=\,\LAP{\etaTs}\!\left(\NABLA{\etaTs} d\alpha\right) 
\end{align}
since $\nabla^{(\etaTs)}\alpha = d\alpha$ for all $0$-forms $\alpha$. Thus, if we define
\begin{align}\label{TJW_DEF_PSIT}
	\psiT \,\equiv\, \NABLA{\etaTs}d\alpha
\end{align}
then from (\ref{TJW_LAP_T}) and (\ref{TJW_LAP}):
\begin{align*}
	\LAP{\etaTs}(\alpha) \,=\, -\delta d\alpha \,=\, 0 \qquad\Longrightarrow\qquad \LAP{\etaTs}\!\left(\NABLA{\etaTs} d\alpha\right)  \,=\, \LAP{\etaTs}\!\left(\psiT\right) \,=\, 0.
\end{align*}
Furthermore, directly from (\ref{TJW_DEF_DIV}):
\begin{align}\label{TJW_DIV_LAP_EQ}
	\DIV{\etaTs}(\nabla^{(\etaTs)}\beta) \,=\, \LAP{\etaTs}(\beta) \qquad \FORALL\quad \beta\in\Gamma\Lambda^{1}\M
\end{align}
and so with $\beta = d\alpha$, it follows from (\ref{TJW_DEF_PSIT}), (\ref{TJW_DIV_LAP_EQ}), (\ref{LAP_COMM_REL}) and (\ref{TJW_LAP_ALP}) respectively: 
\begin{align*}
	\DIV{\etaTs}(\psiT)\,=\,\DIV{\etaTs}(\nabla^{(\etaTs)}d\alpha) \,=\, \LAP{\etaTs}(d\alpha) \,=\, d(\LAP{\etaTs}(\alpha)) \,=\, -d(\delta d\alpha),
\end{align*}
which vanishes when $\delta d\alpha=0$. Hence $\LAP{\etaTs}(\psiT)=0$ and $\DIV{\etaTs}(\psiT)=0$ are both satisfied for {\it any} suitably differentiable complex scalar field $\alpha$ satisfying $\delta d \alpha=0$. This should be compared with (\ref{BOX}), the equation determining the class of pulsed electromagnetic field solutions discussed in the previous section.\\

From the Levi-Civita differential operator (\ref{TJW_LEVI_DIFF_OP}), for any scalar field $\alpha$:
\begin{align*}
	\psiT \,=\, \NABLA{\etaTs}d\alpha \,=\, e^{a} \tensor e^{b}\left[ \,\nabla_{X_{a}}\nabla_{X_{c}}\alpha - \nabla_{\nabla_{X_{a}X_{c}}}\alpha\,\right]
\end{align*}
which describes a symmetric, second-rank covariant tensor. By definition:
\begin{align*}
	 \TR{\etaTs}(\psiT) \,=\, \TR{\etaTs}\!\left(\,\NABLA{\etaTs}d\alpha\,\right) &\,=\, (\NABLA{\etaTs}d\alpha)(X_{i},X^{i}) \,=\, - \LAP{\etaTs}(\alpha) \,=\, \delta d\alpha
\end{align*}
using (\ref{TJW_LAP_T}) and hence $\psiT=\NABLA{\etaTs}d\alpha$ is traceless if $\delta d \alpha=0$.\\

Thus, we conclude that any complex scalar field $\alpha$ satisfying $\delta d\alpha =0$ will give rise to the real metric: 
\begin{align*}
	\g &\,=\, \etaT + \psiT' \,=\, \etaT + \mathrm{Re}(\psiT) \,=\, \etaT + \mathrm{Re}(\,\NABLA{\etaTs}d\alpha\,)
\end{align*}	
generating a solution to the {\it linearised source-free Einstein equation on the perturbative domain} $\UUP$ (\ref{LIN_EIN_SF}).\\

In a local chart $\Phi_{\mathcal{U}}$ possessing dimensionless coordinates $\{t,\rho,\phi,z\}$ with $t\geq 0$, $\rho>0$,
$\phi\in{[0,2\pi)}$ and $|z|\geq 0 $ on a spacetime domain $\mathcal{U}\subset \M$, a local coframe $\CF$ adapted to these co-ordinates is $\{e^{0}=dt,\, e^{1}=d\rho, \,e^{2}=\rho\,d\phi, \,e^{3}=dz\}$.  With $\etaT$ given by (\ref{ETA}), this coframe is $\etaT$-orthonormal but not in general $\g$-orthonormal. Such a chart facilitates the coordination of a series of massive test particles initially arranged in a series of concentric rings with different values of $\rho$ lying in spatial planes with different values of $z$ at $t=0$. Furthermore, we define for any  metric $\g=\etaT+\h$ on $\UU$:
\begin{align*}
	\HMAX(t,\rho,\phi,z) &\,\equiv\, \mathop{\text{max}}_{0\leq a,b\leq 3} \,\left\vert\df\h(X_{a},X_{b})\right\vert.
\end{align*}
The particular complex scalar $\alpha$ of relevance here satisfying $\LAP{\etaTs}(\alpha)=0$ is given in the $(t,\rho,\phi,z)$ chart $\Phi_{\UU}$ above as\footnote{%
	This should be compared with (\ref{alp}).}
\begin{align}\label{CYL_ALP}
	\alpha(t,\rho,z) \,=\, \frac{\W^{2}}{\rho^{2}+[\,a+i(z-t)\,]\,[\,b-i(z+t)\,]} 
\end{align}
with strictly positive real dimensionless constants. The scalar $\alpha(t,\rho,z)$ is then singularity-free in $t$, $\rho$ and $z$ and clearly axially-symmetric with respect to rotations about the $z$-axis. It also gives rise to an {\it axially-symmetric} complex tensor $\psiT^{(0)}$ satisfying\footnote{%
	Since $\NABLA{\etaTs}$ is a flat connection, if $K$ is an $\etaT$-Killing vector then the operator $\NABLA{\etaTs}\Lie_{K} = \Lie_{K}\NABLA{\etaTs}$  on all tensors.%
	} %
$\Lie_{\partial/\partial\phi}\psiT^{(0)}=0$. In $\UU$, this yields the real axially-symmetric metric tensor $\g^{(0)}$ (satisfying $\Lie_{\partial/\partial\phi}\,\g^{(0)}=0$) with
\begin{align*}
	\left\{	\begin{array}{rlcrlcrl}
				\g^{(0)}_{00} &\!=\, \ds -1 + \frac{\partial^{2}\alpha^{\prime}}{\partial t^{2}}, & & \g^{(0)}_{01} &\!=\, \ds \g^{(0)}_{10} \,=\, \frac{\partial^{2}\alpha^{\prime}}{\partial t\,\partial \rho}, & &	\g^{(0)}_{03} &\!=\, \ds \g^{(0)}_{30} \,=\, \frac{\partial^{2}\alpha^{\prime}}{\partial t\,\partial z}  \\[0.5cm]
				\g^{(0)}_{11} &\!=\, \ds 1+\frac{\partial^{2}\alpha^{\prime}}{\partial \rho^{2}}, & & \g^{(0)}_{13} &\!=\, \ds \g^{(0)}_{31} \,=\, \frac{\partial^{2}\alpha^{\prime}}{\partial \rho\,\partial z} &&&\\[0.5cm]
				\g^{(0)}_{22} &\!=\, \ds 1 + \frac{1}{\rho}\frac{\partial\alpha^{\prime}}{\partial \rho}, & & \g^{(0)}_{33} &\!=\, \ds 1 + \frac{\partial^{2}\alpha^{\prime}}{\partial z^{2}} &&&
			\end{array}\right.
\end{align*}
where $\alpha^{\prime}\equiv\mathrm{Re}(\alpha)$.\\

\RRZP{0.9}{%
	This figure illustrates the nature of the spacetime geometry determined by $\g_{0}$ on $\M$ with associated Ricci curvature scalar $\CVRR{0}(t,\rho,z)$. Regions where $\CVRR{0}(t,1,z)$ change sign are clearly visible in the right side where a $2$-dimensional density plot shows a pair of prominent loci that separately approach the future ($t\geq 0$) light-cone of the event at $\{\rho=1,\,t=0,\,z=0\}$. A more detailed graphical description of $\CVRR{0}(t,1,z)$ is given in the left hand $3$-dimensional plot where an initial pulse-like maximum around $t\simeq 0$ evolves into a pair of enhanced loci with peaks at values of $z$ with {\it opposite signs} when $t\geq 1$. This Ricci curvature scalar is generated from a metric perturbation pulse with parameters $a=b=1$ and $\W=1/4$.%
	}

Complex symmetric tensors $\psiT^{(m)}$ with integer chirality $m>0$ satisfying $\LAP{\etaTs}(\psiT^{(m)})=0$, $\DIV{\etaTs}(\psiT^{(m)})=0$, $\TR{\etaTs}(\psiT^{(m)})=0$ and $\frac{1}{i}\Lie_{\partial/\partial\phi}\psiT^{(m)}=m \psiT^{(m)}$  may be generated from $\psiT^{(0)}$ by repeated covariant differentiation with respect to a particular $\etaT$-null, $\etaT$-Killing complex vector field $K_{+}$:
\begin{align*}
	\psiT^{(m)} \,=\, \underbrace{\NABLA{\etaTs}_{K_{+}}\cdots\cdots\NABLA{\etaTs}_{K_{+}}}_{\text{$m$ times}}\psiT^{(0)} 
\end{align*}
where 
\begin{align*}
	K_{+} \,\equiv\, \etaT^{\inv}\!\left(\,d(\,\rho e^{i\phi}\,),-\right)\,=\, e^{i\phi}\left(\frac{\partial}{\partial \rho} + \frac{i}{\rho}\frac{\partial}{\partial\phi}\right).
\end{align*}
Solutions with negative integer chirality can be obtained by complex conjugation of the positive chirality complex eigen-solutions. Each $\psiT^{(m)}$ defines a real spacetime metric $\g^{(m)}=\etaT + \text{Re}(\psiT^{(m)})$ on $\mathcal{U}$ which, for $m\neq 0$, is not axially symmetric:  $\Lie_{\partial/\partial\phi}\g^{(m)}\neq 0$.\\

A longstanding problem in astrophysics has been to understand the origin of the extensive jets that have been observed in X-ray spectroscopy emanating from a number of galaxies. These are considered to be the result of matter falling onto accretion disks that initiate complex magnetohydrodynamic processes. However, it is unclear whether such processes are sufficient to account for the energetics involved. Another puzzle is why some galaxies are observed with bi-directional jets whereas others are not. It has been suggested that gravitation may also play a role in initiating such processes \cite{mashhoonJets}. We have suggested in \cite{TW_Chiral,TW_GPulses} that an analysis of the timelike geodesics associated with a propagating, gravitational chiral pulse may provide some insight into the relevance of gravitational chirality in initiating the magnetohydrodynamic mechanism. As we shall show, chiral pulses can be shown to account for both uni-directional and bi-directional jet structures. \\

Given $\psiT$ and hence $\g=\etaT + \h$, all dimensionless proper-time parametrised timelike spacetime geodesics $C$ on $\mathcal{U}$ (with tangent vector $\dot{C}$) associated with $\g$  must satisfy a differential-algebraic system. For any worldline $C$ with components $C^{\mu}(\tau)$ in any local chart on $\mathcal{U}$ with dimensionless coordinates $\{x^{\mu}\}$  and  $\dot{C}^{\mu}(\tau)=\partial C^{\mu}(\tau)/\partial\tau$:
\begin{align}
	\begin{split}\label{DAE}
		\left.\NABLA{\g}_{\dot{C}(\tau)}\dot{C}(\tau)\right|_{C(\tau)} &\,=\, 0 \\[0.2cm]
		\left.\g(\,\dot{C}(\tau),\dot{C}(\tau)\,)\right|_{C(\tau)} &\,=\, -1.
	\end{split}
\end{align}	
i.e.:
\begin{align*}
	\left.\NABLA{\g}_{\dot{C}(\tau)}\dot{C}(\tau)\right|_{C(\tau)} &\,\equiv\, \left.\frac{D\dot{C}^{\mu}(\tau)}{d\tau} \,\frac{\partial}{\partial x^{\mu}}\right|_{C(\tau)} \,=\, \left.\left( \frac{d\dot{C}^{\mu}(\tau)}{d\tau} +(\Gamma^{\mu}_{\alpha\beta} \circ C)(\tau) \,\dot{C}^{\alpha}(\tau)\,\dot{C}^{\beta}(\tau)\right)\frac{\partial}{\partial x^{\mu}}\right|_{C(\tau)}
\end{align*}
where $\Gamma^{\mu}_{\alpha\beta}$ denotes a Christoffel symbol associated with $\NABLA{\g}$.\\

In the following only solutions to (\ref{DAE}) that lie in the perturbative domains $\UUP$, as determined by (\ref{perturb_domains}), are displayed. \\

\HMAXRRZP{0.9}{%
	The axially-symmetric expressions $\vert\CVRR{0}(t,\rho,0)\vert$ and $\HMAX(t,\rho,0,0)$ are plotted as functions of $\rho$ for $t=0,\,0.25,\,0.5,\,0.75$ and parameters $a=b=1$, $\W=1/4$. Regions where the blue curves lie under the red dotted line denote perturbative regions $\mathcal{P}_{\mathcal{U}}$. The light blue shaded regions clearly indicate curvature scalars that are greater in magnitude than unity despite lying within $\mathcal{P}_{\mathcal{U}}$ regions. 
	}
	
An indication of the nature of the spacetime geometry determined by $\g^{(m)}$ on $\M$  is given by the structure of the associated Ricci curvature scalar $\CVRR{m}(t,\rho,z)$. Unlike gravitational wave spacetimes  this scalar is not identically zero. For $m=0$ it is axially symmetric and in the chart $\Phi_{\UU}$ its independence of  $\phi$  means that  for values of fixed radius $\rho_{0}$  its structure can be displayed for a range of  $t$ and $z$  values given a choice of parameters $(a,b,\W)$. Regions where $\CVRR{0}(t,1,z)$ change sign are clearly visible in the right side of  figure~\ref{fig:rrzp} where a $2$-dimensional density plot shows a pair of prominent loci that separately approach the future ($t\geq0$)  light-cone of the event at $\{\rho=1,t=0,z=0\}$. A more detailed graphical description of $\CVRR{0}(t,1,z)$ is given in the left hand $3$-dimensional plot in figure~\ref{fig:rrzp} where an initial  pulse-like  maximum around $t\simeq 0$ evolves into a pair of enhanced loci with peaks at values of $z$ with {\it opposite signs} when $t\geq 1$. In this presentation the maximum pulse height has been normalised to unity. This characteristic behaviour is similar to that possessed by $\mathrm{Re}(\,\alpha(t,1,z)\,)$. It suggests that ``tidal forces'' (responsible for the geodesic deviation of neighbouring geodesics \cite{schutz,laemmerzahl,perlick}) are concentrated in spacetime regions where components of the Riemann tensor of $\g^{(0)}$ have pulse-like behaviour in domains similar to those  possessed by $\CVRR{0}(t,\rho,z)$.\\

\ZPZLAYERS{0.7}{%
	On the left  six geodesics are shown emanating from  six locations with $\phi$ values $0, \pi/3, 2\pi/3, \pi, 4\pi/3, 5\pi/3  $  on a  ring  with radius $10^{-4}$ in the plane $z=0.735$ and six from similarly arranged points on rings of the same radius at  $z=0.245$, $z=-0.245$ and $z=-0.735$. The initial locations are not resolved in these figures. The 24 geodesics each evolve from $\tau=0$ to $\tau=10^{4}$ and clearly display an axially symmetric  bi-directional jet structure from the rings in conformity  with the expectations based on the spacetime structure of $\CVRR{0}(t,1,z)$ in figure~\ref{fig:rrzp}. The figure on the right resolves the structure of this jet array for $0\leq\tau\leq 100$. All geodesics are generated with the additional initial  conditions $\dot{\rho}(0) = 0$, $\dot{z}(0) = 0$, $\dot{\phi}(0) = 0.4$ and the background perturbation pulse has parameters $a=b=1$, $\W = 1/6$.  A single uni-directional jet-array arises when only one ring is populated with matter. This figure demonstrates that the jets from the sources at $z=\pm 0.245$ have a dimensionless aspect ratio $\mathcal{A}(10^{4})=64.7$ much greater than those produced from the sources at $z=\pm 0.735$ where $\mathcal{A}(10^{4})=3.22$. %
	}

Explicit formulae for $\CVRR{0}(t,\rho,z)$ and $\HMAX(t,\rho,\phi,z)$ are not particularly illuminating\footnote{%
	Since $\g^{(0)}$ is axially-symmetric, the function $\HMAX$ is independent of $\phi$. 
	}. %
However, for fixed values of the parameters $(a,b,\W)$, their values can be plotted numerically in order to gain some insight into their relative magnitudes in any perturbative domain $\mathcal{P}_{\mathcal{U}}$. With $z$ fixed at zero, figure~\ref{fig:hmaxrrzp} displays such plots as functions of $\rho$ and a set of $t$ values. It is clear that in perturbative domains the curvature scalar may exceed unity. Since in general:
\begin{align*}
	\CVRR{0}(t,\rho,z) \,=\, \mathcal{Q}(t,\rho,z)\kappa^{2} + O(\kappa^{3})
\end{align*}	
where $\mathcal{Q}$ is a non-singular rational function of its arguments and the tensor $\texttt{h}$ is, by definition, of order $\kappa$, figure~\ref{fig:hmaxrrzp} demonstrates that relative tensor $\kappa$-orders are not, in general, indicators of their corresponding relative magnitudes. \\

\ZPZLAYERSQ{0.8}{%
	On the left, six geodesics are shown emanating from  six locations with $\phi$ values $0, \pi/3, 2\pi/3, \pi, 4\pi/3, 5\pi/3  $  on a  ring  with radius $10^{-4}$ in the plane $z=0.735$ and six from similarly arranged points on rings of the same radius at $z=-0.735$. The initial locations are not resolved in these figures. The $12$ geodesics each evolve from $\tau=0$ to $\tau=10^{4}$ and clearly display an axially symmetric uni-directional jet structure from the rings. All geodesics are generated with the additional initial  conditions $\dot{\rho}(0) = 0$, $\dot{z}(0) = 0$, $\dot{\phi}(0) = 0.4$ and the background perturbation pulse has parameters $a=1$, $b=3$, $\W= 1/6$. The figure in the centre shows an oppositely directed jet evolving from similar initial conditions, but with initial $z=0.245$ and $z=-0.245$. The figure on the right displays the ``$z$-asymmetric'' jet structure obtained by merging both pairs of sources with {\it dominant} component belonging to the jet in the left-hand figure having aspect ratio $\mathcal{A}(10^{4})=178.8$.%
	}

By modelling thick accretion disks by a finite number of massive point particles occupying a number of planar rings with varying $\rho$ and varying $z$ the system (\ref{DAE}) has been explored numerically. Sets of  space-curves in spacelike sections of the perturbative spacetimes defined above are displayed in figure~\ref{fig:zpzlayers} and figure~\ref{fig:zpzlayersq} for various choices of the dimensionless parameters $a,b$, $\W$ and dimensionless evolution proper-time. The resulting jet-like structures of these space-curves with dimensionless proper-time parameter $\tau\in[0,\taumax]$ are quantified in terms of aspect ratios defined by:
\begin{align*}
	\AR &\,\equiv\, \left| \cfrac{\wh{z}(\taumax)-\wh{z}(0)}{\wh{\rho}(\taumax)-\wh{\rho}(0)} \right|.
\end{align*}

\section{Spinor Pre-potentials}\label{sect:SPINOR}
In the above sections we have outlined how a particular $0$-form singularity-free, dispersive pulse-like solution $\alpha$ on spacetime for the scalar equation (\ref{BOX}) in a Minkowski background can be used to generate solutions to {\it linear} partial differential equations of relevance to mathematical physics, namely the Minkowski source-free vacuum Maxwell system for electromagnetic fields, the linear system arising from the Minkowski Bopp-Land\'{e}-Podolsky source-free vacuum system and the linearised equation of source-free Einstein theory of gravitation perturbed about Minkowski spacetime. A notable feature of all these solutions is that they include non-stationary, non-separable functions when expressed in local coordinates. Furthermore they can all be classified in terms of chirality eigenvalues of certain differential operators with respect to Minkowski Killing vector fields. \\

This raises the question of whether there exist classical linear spinor equations that exhibit similar spinor pulse-like solutions that can be generated from the particular complex pulse-like solutions to (\ref{BOX}). To show that this is indeed the case it proves expedient to exploit the close relation of spinor representations of the covering of the $\g$-orthonormal group $\text{SO}(3,1)$ with the spacetime Clifford algebra $\text{C}_{3,1}(V,\GG)$ where $V=\Gamma\Lambda^{1}\M$ is the underlying vector space with cobasis $\{e^{0},e^{1},e^{2},e^{3}\}$ and $\GG$ has signature $(-,+,+,+)$.\\

In order to maintain the dimensional coherence of equations that relate physical observables, it is convenient to take all the elements in $V$ to be physically dimensionless so that all elements in the Clifford algebra $\text{C}_{3,1}(V,\GG)$ are also physically dimensionless. This can be readily achieved if all local charts for Minkowski spacetime are defined with dimensionless coordinate labels. In which case, $\GG\equiv\g$ is dimensionless as in the sections above. In scenarios where one chooses to assign any other physical dimension to the metric tensor $\g$ which takes the value $\Lambda_{0}^{2}$ in some system of physical units (i.e. $\PHYSDIM{\g}=\Lambda_{0}^{2}$) then $\GG\equiv\g/\Lambda^{2}_{0}$ ensures that $\GG$ is physically dimensionless with signature $(3,1)$ and each element in $V$ is dimensionless. In this section, the Levi-Civita connection associated with the metric tensor $\GG$ will be denoted $\nabla$, as elsewhere. \\

We denote by $\Gamma\Lambda_{3,1}\M$ the $16$-dimensional vector space with algebra product $\w$ where an arbitrary element of $\Gamma\Lambda_{3,1}\M$ is a superposition of all $p$-forms ($p=0,1,2,3$). If the coefficients of each $p$-form in this superposition belong to $\RR$ (or $\CC$) the elements are said to be {\it over the field} $\RR$ (or $\CC$). The space $\Gamma\Lambda_{3,1}\M$ is a space of local sections of the exterior algebra bundle over Minkowski spacetime $\M$ and we call its elements {\it inhomogeneous forms} when projected to $\M$ by pullback. Homogeneous elements of $\Gamma\Lambda_{3,1}\M$ are then eigen-forms of the involution $\eta$ where 
\begin{align*}
	\eta\,\Phi \,=\, (-1)^{p}\,\Phi \qquad\Phi\in\Gamma\Lambda^{p}\M \quad (p=0,1,2,3).
\end{align*}
Suppose $\sigma$ is a $1$-form in $\Gamma\Lambda_{3,1}\M$ and $\Omega$ an arbitrary element in $\Gamma\Lambda_{3,1}\M$, then one defines a new algebra $\text{C}_{3,1}(V,\GG)$ with underlying vector space $V$ and algebra product $\vee$ by the rules:
\begin{align}
	\label{RWT_HASH1}	\sigma \vee \Omega &\,=\, \sigma \w \Omega + i_{\wt{\sigma}}\Omega \\[0.2cm] 
	\label{RWT_HASH2}   \Omega \vee \sigma &\,=\, \sigma \w \eta\,\Omega - i_{\wt{\sigma}}(\,\eta\,\Omega\,)
\end{align}
where the tilde map is now taken with respect to $\GG$. The associativity of this Clifford product completely determines the action of $\vee$ between arbitrary forms and the vector space of exterior forms is thereby associated with a Clifford algebra. Thus, if $\{e^{a}\}$ and $\{X_{b}\}$ are any dual bases of $\Gamma T^{*}\M$ and $\Gamma T\M$ respectively, and $\Omega_{1},\Omega_{2}$ any exterior forms on $\M$, the relations:
\begin{align}
	\label{RWT_E1}	\Omega_{1}\vee\Omega_{2} &\,=\, \sum_{p=0}^{4}  \frac{(-1)^{\lfloor p/2\rfloor}}{p!}\,(\eta^{p}\, i_{X_{a_{1}}}i_{X_{a_{2}}} \cdots i_{X_{a_{p}}} \,\Omega_{1}) \w ( i_{\wt{e^{a_{1}}}}i_{\wt{e^{a_{2}}}} \cdots i_{\wt{e^{a_{p}}}}\,\Omega_{2}) \\[0.2cm]
	\label{RWT_E2}	\Omega_{1}\w\Omega_{2} &\,=\,  \sum_{p=0}^{4} \frac{(-1)^{\lfloor p/2\rfloor}}{p!}\,(i_{X_{a_{1}}}i_{X_{a_{2}}} \cdots i_{X_{a_{p}}} \,\eta^{p}\,\Omega_{1}) \vee ( i_{\wt{e^{a_{1}}}}i_{\wt{e^{a_{2}}}} \cdots i_{\wt{e^{a_{p}}}}\,\Omega_{2})
\end{align}
follow recursively from (\ref{RWT_HASH1}) and (\ref{RWT_HASH2}). In these relations the notation $\lfloor r \rfloor$ denotes the integer part of $r$.  If one takes $\Omega\equiv e^{a}$ and $\sigma\equiv e^{b}$ in (\ref{RWT_HASH1}), (\ref{RWT_HASH2}) one has:
\begin{align*}
	e^{b} \vee e^{a} &\,=\, e^{b} \w e^{a} + \GG^{\inv}(e^{a},e^{b}) \\[0.2cm]
	e^{a} \vee e^{b} &\,=\, -e^{b} \w e^{a} + \GG^{\inv}(e^{a},e^{b})
\end{align*}
or 
\begin{align*}
	e^{b} \vee e^{a} + e^{a} \vee e^{b} \,=\, 2\GG^{\inv}(e^{a},e^{b})
\end{align*}
in terms of the metric tensor $\GG^{\inv}$. We denote sections of the Clifford bundle over $\M$ by $\Gamma\text{C}_{3,1}(V,\GG)\M$. {\it In particular}, if $\{e^{a}\}$ is a {\it local} $\GG$-{\it orthonormal basis} (\ref{RWT_HASH1}) and (\ref{RWT_HASH2}) yield:
\begin{align*}
	[e^{b}\vee e^{c},\, e^{a}] \,=\, 2(\etaT^{ac}\,e^{b} - \etaT^{ab}\,e^{c})
\end{align*}
where $\etaT^{ab}\equiv \GG^{\inv}(e^{a},e^{b})$ and
\begin{align}\label{CliffCOMM}
	[\Omega_{1},\,\Omega_{2}] \,\equiv\, \Omega_{1}\vee\Omega_{2} - \Omega_{2}\vee\Omega_{1}
\end{align}
denotes the {\it Clifford commutator} between arbitrary elements in the Clifford algebra. If the Levi-Civita connection $1$-forms are denoted $\{\omega_{bc}\}$ in this cobasis, one defines for any local vector field $Y\in\Gamma T\M$:
\begin{align}\label{RWT_ADD}
	\Sigma_{Y}\,\equiv\, \frac{1}{4}\,\omega_{bc}(Y)\,e^{b} \vee e^{c}
\end{align}
so that the relation 
\begin{align*}
	\nabla_{X_{a}}e^{c} \,=\, -\omega^{c}{}_{b}(X_{a})\,e^{b}
\end{align*}
yields the identity
\begin{align}\label{RWT_STAR}
	\nabla_{Y}e^{a} \,\equiv\, [\Sigma_{Y},e^{a}], \qquad a=0,1,2,3 \QUAD{and} Y\in\Gamma T\M
\end{align}
in terms of the Clifford commutator. Since the exterior product of {\it mutually} $\GG${\it -orthogonal} $1$-forms is the same as their Clifford product then (\ref{RWT_STAR}) extends to:
\begin{align*}
	\nabla_{Y}e^{I} \,=\, [\Sigma_{Y},e^{I}]
\end{align*}
where $\{e^{I}\}$ denotes a $\GG$-orthonormal Clifford product multi-basis for $\Gamma\text{C}_{3,1}(V,\GG)\M$. \\

The Clifford algebra $\text{C}_{3,1}(V,\GG)$ can be decomposed into four {\it minimal left ideals} characterised by a complete set of four minimal rank (primitive) idempotents $\{\proj_{1},\proj_{2},\proj_{3},\proj_{4}\}$ that are:
\begin{alignat*}{2}
	\text{``pairwise orthogonal'':}&\quad\proj_{i}\vee \proj_{j} \,=\, 0 	&&\qquad\text{(for $i\neq j$)}\\
	\text{and idempotent:}&\quad \proj_{i}^{2}\,=\, \proj_{i} 				&&\qquad\text{($i=1,2,3,4$)} 
\end{alignat*}
with
\begin{align*}
	\sum_{j=1}^{4}\proj_{j} \,=\, 1. 
\end{align*}
The set is said to be {\it maximal} when it is not possible to find a set of more idempotents having these properties. 
The {\it primitivity} of each $\proj_{j}$ implies
\begin{align*}
	\proj_{j} \vee \Omega \vee \proj_{j} \,=\, 4\,S_{0}(\Omega\vee\proj_{j})\,\proj_{j} \qquad\text{for all $\proj_{j}, \Omega\in\Gamma\text{C}_{3,1}(V,\GG)\M$}
\end{align*}
where the map $S_{0}$ extracts the $0$-form part of the Clifford product $\Omega\vee\proj_{j}$ using (\ref{RWT_E1}).\\

For example, a particular set of idempotents for $\text{C}_{3,1}(V,\GG)$ is given by:
\begin{align}\label{P_idem}
	\left\{	\begin{array}{rl}
				\proj_{1}	&\ds\!\!\!\equiv\;	\frac{1}{4}\,( 1 + e^{1}) \vee (1 + e^{0} \vee e^{2}) \\[0.4cm]
				\proj_{2}	&\ds\!\!\!\equiv\;	\frac{1}{4}\,( 1 + e^{1}) \vee (1 - e^{0} \vee e^{2}) \\[0.4cm]
				\proj_{3}	&\ds\!\!\!\equiv\;	\frac{1}{4}\,( 1 - e^{1}) \vee (1 + e^{0} \vee e^{2}) \\[0.4cm]
				\proj_{4}	&\ds\!\!\!\equiv\;	\frac{1}{4}\,( 1 - e^{1}) \vee (1 - e^{0} \vee e^{2}).
			\end{array}\right.
\end{align}
Since $e^{0},e^{3}$ and $e^{0}\vee e^{3}$ all have Clifford squares equal to $\pm 1$, they all have Clifford inverses $(e^{0})^{-1}$, $(e^{3})^{-1}$ and $(e^{0} \vee e^{3})^{-1}$ respectively. Thus the above four primitive idempotents are all {\it  similar} under the transformations: 
\begin{align}\label{P_trans}
	\left\{	\begin{array}{rl}
				e^{3}\vee \proj_{1}\vee(e^{3})^{-1} &\!\!=\; \proj_{3} \\[0.4cm]
				e^{0}\vee \proj_{1}\vee(e^{0})^{-1} &\!\!=\; \proj_{4} \\[0.4cm]
				(e^{0} \vee e^{3})\vee\proj_{1}\vee (e^{0} \vee e^{3})^{-1} &\!\!=\; \proj_{2}.
			\end{array}\right.
\end{align}
Since $\text{C}_{3,1}(V,\GG)$ over $\RR$ or $\CC$ is isomorphic to a total matrix algebra, it is always possible to construct a basis $\mbasis{ij}\in\Gamma\text{C}_{3,1}(V,\GG)\M$ ($i,j=1,2,3,4$) for it, satisfying: 
\begin{alignat*}{2}
	\mbasis{ij} \vee \mbasis{jk} &\,=\, \mbasis{ik}	\qquad 	&& (\text{no sum}) \\[0.2cm]
	\mbasis{ij} \vee \mbasis{pk} &\,=\, 0 			\qquad	&& j\neq p, \, k=1,2,3,4.
\end{alignat*}
There exists a whole class of such bases related by the {\it inner automorphism}:
\begin{align*}
	\mbasis{ij} \,\longmapsto\, s \vee \mbasis{ij} \vee s^{-1} 
\end{align*}
where $s$ is any invertible element in $\Gamma\text{C}_{3,1}(V,\GG)\M$. Elements of this class constitute a matrix basis for $\Gamma\text{C}_{3,1}(V,\GG)\M$ since if we write $\phi=\sum_{i,j}\phi_{ij}\,\mbasis{ij}$, $\psi=\sum_{i,j}\psi_{ij}\,\mbasis{ij}$ where $\phi_{ij},\psi_{ij}$ are complex functions on $\M$ then:
\begin{align*}
	\phi \vee \psi \,=\, \sum_{i,j} \rho_{ij}\,\mbasis{ij}
\end{align*}
where $\rho_{ij}\equiv \sum_{k=1}^{4}\phi_{ik}\psi_{kj}$. In terms of $\{\proj_{j}\}$, the basis elements $\{\mbasis{ij}\}$ satisfy:
\begin{align*}
	\mbasis{ij} \,=\, \proj_{i} \vee c_{ij}\vee\proj_{j} \,=\, \proj_{i} \vee c_{ij} \,=\, c_{ij} \vee \proj_{j} \qquad (\text{no sum})
\end{align*}
for some Clifford sections $c_{ij}\in\Gamma\text{C}_{3,1}(V,\GG)\M$. \\

Furthermore, the relations (\ref{P_trans}) enable one to readily construct a real $(4\times 4)$-matrix (Majorana) representation for any real element in $\Gamma\text{C}_{3,1}(V,\GG)\M$ \cite[p. 17]{benntucker}.\\

Any local section $\Psi\in\Gamma\text{C}_{3,1}(V,\GG)\M$ belongs to a {\it minimal left ideal} denoted $\Gamma\mathcal{I}_{L}\proj\M$ represented by a local primitive idempotent $\proj\in\Gamma\text{C}_{3,1}(V,\GG)\M$ if $\Psi=\Psi\vee \proj$. Such a vector space carries a representation of the groups $\text{PIN}(3,1)$ and $\text{SPIN}(3,1)$. As such, the elements $\Psi$ are identified as local {\it spinor field representations}. In Minkowski spacetime $\M$ they can be matched across overlapping charts on $\M$ to generate global spinor representations of these spin groups. \\

Recall that the (torsion-free) Levi-Civita covariant derivative is defined to be a type-preserving operator on sections of the bundle of tensors over spacetime $\M$ satisfying the Leibniz rule (over $\tensor$ and hence $\w$):
\begin{align*}
	\nabla_{Y}(\,\alpha \tensor \beta\,) \,=\, \nabla_{Y}\alpha \tensor \beta + \alpha \tensor \nabla_{Y}\beta \qquad\text{for all  } Y\in\Gamma T\M, \,\alpha\in\Gamma T^{r_{1}}_{s_{1}}\M, \,\beta\in\Gamma T^{r_{2}}_{s_{2}}\M
\end{align*}
that is also compatible with the spacetime metric-tensor field $\g$:
\begin{align*}
	Y\left(\df\g(Z_{1},Z_{2})\,\right) \,=\, \g(\nabla_{Y}Z_{1},Z_{2}) + \g(Z_{1},\nabla_{Y}Z_{2}) \qquad\text{for all  } Y, Z_{1},Z_{2}\in\Gamma T\M.
\end{align*}
The construction of covariant derivatives of spinor fields differs fundamentally from the construction of the Levi-Civita covariant derivative of tensor fields in a number of ways. For the bundle $\Gamma\text{C}_{3,1}(V,\GG)\M$ spinor covariant  derivatives may be classified in terms of the {\it involution operators} $\{\xi,\xi\eta,\xi^{*},\xi\eta^{*}\}$ on elements of the Clifford bundle. For any such Clifford sections $\Omega_{1},\Omega_{2}$:
\begin{align*}
	\eta\!\left(\df\Omega_{1} \vee \Omega_{2}\,\right) &\,=\, \eta\,\Omega_{1} \vee \eta\,\Omega_{2} \\
	\xi\!\left(\df\Omega_{1} \vee \Omega_{2}\,\right) &\,=\, \xi\,\Omega_{2} \vee \xi\,\Omega_{1} \\
	\left(\df\Omega_{1} \vee \Omega_{2}\,\right)^{\!*} &\,=\, \Omega_{1}^{*} \vee \Omega_{2}^{*}. 
\end{align*}
In these rules, $*$ denotes complex conjugation and:
\begin{align*}
	\xi\Omega \,=\, \xi\sum_{r=0}^{4}\Omega_{r} \,=\, \sum_{r=0}^{4}\xi\Omega_{r} \,=\, \sum_{r=0}^{4}(-1)^{\lfloor r/2\rfloor}\Omega_{r},
\end{align*}
where $\Omega_{r}\in\Gamma\Lambda^{r}\M$ when $\Omega$ is expressed as a superposition of degree $r$-forms in\footnote{Equivalently, $\xi$ maps any scalar or element of $V$ to itself but reverses the order of elements in any Clifford product.} $\Gamma\Lambda\M$. An alternative notation for $\xi\Omega$ that we shall use below is $\Omega^{\xi}$. Furthermore, this notation will be adopted for any of the involutions given above. If $J$ denotes any one of these four involutions, the {\it adjoint of any spinor field} $\Psi\in\Gamma\mathcal{I}_{L}\proj\M$ is defined as:
\begin{align*}
	\Psi^{\dagger} \,\equiv\, W\vee \Psi^{J}
\end{align*}
for some chosen element $W\in\proj\vee\Gamma\text{C}_{3,1}(V,\GG)\M \vee \proj^{J}$ such that 
\begin{align}\label{RWT_KEY}
	\left( \Omega \vee \Psi \right)^{\dagger} \,=\, \Psi^{\dagger} \vee \Omega^{J} \qquad\text{for all  } \Omega\in\Gamma\text{C}_{3,1}(V,\GG)\M.
\end{align}
The equation (\ref{RWT_KEY}) in turn defines an ``inner product'' or {\it spinor bilinear form} $(\;,\;)$ on the space of spinor sections generated by $\proj$ and $W$:
\begin{align*}
	(\phi,\psi) \QQUAD{with} \phi,\psi\in\Gamma\mathcal{I}_{L}\proj\M  
\end{align*}
such that 
\begin{align*}
	(\phi,\psi)\proj \,=\, \phi^{\dagger} \vee \psi.
\end{align*}
Note that this inner product depends upon both $\proj$, the element $W$ and the choice of involution $J$. \\

One may now define a spinor-type-preserving {\it spinor covariant derivative} $S_{Y}$ with respect to any vector field $Y$ on spacetime (i.e. if $\Psi=\Psi\vee\proj$, $S_{Y}\Psi=S_{Y}\Psi\vee\proj$ for all $Y$), satisfies a Leibniz rule over Clifford products:
\begin{align*}
	S_{Y}(\,\Omega \vee \Psi\,) \,=\, \nabla_{Y}\Omega \vee \Psi + \Omega \vee S_{Y}\Psi \qquad\text{for all  }\Omega\in\Gamma\text{C}_{3,1}(V,\GG)\M,\,\Psi\in\Gamma\mathcal{I}_{L}\proj\M
\end{align*}
and is compatible with the chosen spinor adjoint:
\begin{align*}
	Y\,(\Phi,\Psi) \,=\, (S_{Y}\Phi,\Psi) + (\Phi,S_{Y}\Psi).
\end{align*}
In general, these last two conditions do not completely fix $S_{Y}$ for any given $\proj,W,J$. However, for our purpose here, we shall choose $J=\xi^{*}$ in which case the definition 
\begin{align}\label{SY}
	\begin{split}
		S_{Y}:\Gamma\mathcal{I}_{L}\proj\M &\,\longrightarrow\,\Gamma\mathcal{I}_{L}\proj\M \\
		\Psi &\,\longmapsto\, S_{Y}\Psi \,=\, \nabla_{Y}\Psi \vee \proj + (i_{Y}\lambda)\Psi
	\end{split}
\end{align}
uniquely defines $S_{Y}$ when the $1$-form $\lambda$ satisfies:
\begin{align*}
	(i_{Y}\lambda) \vee W \,=\, \frac{1}{2}\proj \vee \nabla_{Y}W \vee \proj^{J}.
\end{align*}
If follows that $S_{Y}$ also depends upon $\proj,W$ and $J\equiv\xi^{*}$. For other choices of involution, the reader may consult \cite{benntucker,Tucker_Clifford,Tucker_LUCY}. \\

In Minkowski spacetime there exists a global chart with dimensionless coordinates $\{x^{0},x^{1},x^{2},x^{3}\}$ and a globally defined $\nabla${\it -parallel} $\GG$-orthonormal cobasis of $1$-forms $\{e^{a}=dx^{a}\}$ ($a=0,1,2,3$) satisfying $\nabla e^{a}=0$. This implies that the idempotent projectors (\ref{P_idem}) are all $\nabla$-parallel. Using (\ref{RWT_ADD}) and (\ref{RWT_STAR}) the spinor covariant derivative (\ref{SY}) can, in this case, be expressed in terms of $\Sigma_{Y}$ as:
\begin{align*}
	S_{Y}\Psi \,=\, \nabla_{Y}\Psi + \Psi \vee \Sigma_{Y}\qquad \text{for any  }\Psi\in\Gamma\mathcal{I}_{L}\proj_{j}\M\quad (j=1,2,3,4).
\end{align*}
In this $\nabla$-parallel cobasis, all connection $1$-forms vanish: $\om{a}{b}=0$ for all $a,b=0,1,2,3$, and hence $S_{Y}\Psi=\nabla_{Y}\Psi$ for all $Y\in\Gamma T\M$. Therefore, the projectors $\proj_{j}$ are {\it $S$-parallel with respect to the involution} $J=\xi^{*}$: $S_{Y}\proj_{j}=0$ for all $Y\in\Gamma T\M$.  However, in a non-$\nabla$-parallel cobasis $\om{a}{b}\neq 0$ for some $a,b$ and one may exploit (\ref{SY}) for some alternative choice of $\proj,W$ or $J$. For example, if one selects a local cylindrically polar chart with dimensionless coordinates $\{t,\rho,\phi,z\}$ in which a local $\GG$-orthonormal cobasis is:
\begin{align*}
	e^{0} \,=\, dt, \quad e^{1} \,=\, d\rho, \quad e^{2} \,=\, \rho\,d\phi, \quad e^{3} \,=\, dz,
\end{align*}
one finds an adapted projector set $\{\wh{\PROJ}_{i}\}$ ($i=1,2,3,4$) in this chart in which {\it not all projectors are} $\nabla$, and hence $S$, parallel with respect to $J=\xi^{*}$:
\begin{align*}
	\left\{	\begin{array}{rl}
				\wh{\PROJ}_{1}	&\ds\!\!\!\equiv\;	\frac{1}{4}\,( 1 - e^{2})\vee (1 + ie^{1}\vee e^{3}) \\[0.4cm]
				\wh{\PROJ}_{2}	&\ds\!\!\!\equiv\;	\frac{1}{4}\,( 1 + e^{2})\vee (1 + ie^{1}\vee e^{3}) \\[0.4cm]
				\wh{\PROJ}_{3}	&\ds\!\!\!\equiv\;	\frac{1}{4}\,( 1 - e^{2})\vee (1 - ie^{1}\vee e^{3}) \\[0.4cm]
				\wh{\PROJ}_{4}	&\ds\!\!\!\equiv\;	\frac{1}{4}\,( 1 + e^{2})\vee (1 - ie^{1}\vee e^{3}).
			\end{array}\right.
\end{align*}
However, one may find an alternative adapted projector set $\{\PROJ_{i}\}$ ($i=1,2,3,4$) in which all projectors {\it are} $\nabla$ and $S$-parallel with respect to $J=\xi^{*}$:
\begin{align}
	\left\{	\begin{array}{rl}
				\PROJ_{1}	&\ds\!\!\!\equiv\;	\frac{1}{4}\,( 1 + ie^{0})\vee (1 + ie^{1}\vee e^{2}) \\[0.4cm]
				\PROJ_{2}	&\ds\!\!\!\equiv\;	\frac{1}{4}\,( 1 + ie^{0})\vee (1 - ie^{1}\vee e^{2}) \\[0.4cm]
				\PROJ_{3}	&\ds\!\!\!\equiv\;	\frac{1}{4}\,( 1 - ie^{0})\vee (1 - ie^{1}\vee e^{2}) \\[0.4cm]
				\PROJ_{4}	&\ds\!\!\!\equiv\;	\frac{1}{4}\,( 1 - ie^{0})\vee (1 + ie^{1}\vee e^{2}).
			\end{array}\right.
\end{align}
If $\{e^{a}\}$ and $\{X_{b}\}$ are naturally dual bases (not necessarily $\GG$-orthonormal) we define, in general, the {\it Dirac differential operator} on spinor sections in $\Gamma\mathcal{I}_{L}\proj\M$:
\begin{align*}
	\slashed{S} \,\equiv\, e^{a} \vee S_{X_{a}}
\end{align*}	
and for all $\Psi=\Psi\vee\proj\in\Gamma\mathcal{I}_{L}\proj\M$ we call 
\begin{align}\label{RWT_HASH3}
	\slashed{S}\Psi \,=\, m_{0}\Psi, \qquad m_{0}\geq 0
\end{align}
a {\it classical generalised Dirac equation} for an electromagnetically neutral spinor field $\Psi$ with dimensionless mass parameter $m_{0}$. Dirac's discovery ($1928$) of the electron-positron matrix field equation was based on a $U(1)$-coupling to the real Maxwell potential $1$-form $A$ on spacetime. In the context of fields in $\Gamma\mathcal{I}_{L}\proj\M$ this employs the {\it complex} $U(1)$ spinor covariant derivative:
\begin{align*}
	S'_{Y}\Psi \,=\, S_{Y}\Psi + i\,q\,A(Y)\,\Psi
\end{align*}
and his wave equation is a matrix representation of the Clifford bundle equation:
\begin{align}\label{RWT_HASH4}
	\slashed{S}\Psi + i\,q\,e^{a}A(X_{a})\,\Psi - m_{e}\Psi \,=\, 0 \qquad\text{for all $\Psi\in \Gamma\mathcal{I}_{L}\proj\M$, $q,m_{e}\in\RR^{+}$}.
\end{align}
In the quantum interpretation of this equation in Minkowski spacetime, stationary time-periodic amplitudes are promoted to operators in a Fock space that create and annihilate ``fundamental'' electron or positron particles in states with properties obtained from a classical analysis of {\it separable solutions} to (\ref{RWT_HASH4}). Since the recent discovery of {\it massive} neutrinos, there is little room left for the physical interpretation of solutions to (\ref{RWT_HASH3}) with $m_{0}=0$ in the Standard Model of fundamental particle physics. Notwithstanding this observation we assert that the same pre-potential complex pulse solution $\alpha$ that generates Maxwell, Bopp-Land\'{e}-Podolsky and linearised vacuum Einstein solutions does indeed generate exact classical non-stationary, non-separable solutions to the massless electrically neutral Dirac equation (\ref{RWT_HASH3}) in Minkowski spacetime. \\

To illustrate this assertion we first choose the minimal left ideal generated by the idempotent $\proj_{1}$. By analogy with (\ref{PIEQNS}), we next introduce a complex $2$-form $\Pi^{\nu,\chi}$ and construct the $1$-form:
\begin{align*}
	A^{\nu,\chi} \,=\, \star d(\, \alpha\,\Pi^{\nu,\chi}\,)
\end{align*}
with $\nu\in\{1,2\}$, $\chi\in\{-1,0,1\}$, 
\begin{align}\label{SPINOR_PI}
	\left\{	\begin{array}{rl}
				\Pi^{1,\pm1}		&\!=\;	d(x\pm i y) \w d t \,=\, e^{\pm i\phi}\left( d\rho \pm i\rho\,d\phi \right) \w dt ,\\[0.2cm]
				\Pi^{1,0}			&\!=\;	d z \w d t,\\[0.2cm]
				\Pi^{2,\chi }		&\!=\; \star\, \Pi^{1,\chi }
			\end{array}\right.
\end{align}
and define the $2$-form:
\begin{align}\label{Fnuchi}
	F^{\nu,\chi} \,=\, dA^{\nu,\chi}.
\end{align}
In a dimensionless Minkowski chart $\{t,x,y,z\}$ with $\GG$-orthonormal cobasis $\{e^{0}= dt,e^{1}=dx,e^{2}=dy,e^{3}=dz\}$ consider first the chirality-zero $\Pi^{1,0}=dz \w dt \equiv dz\vee dt$. The associated complex $2$-form for this choice of $\Pi^{\nu,\chi}$ is then:
\begin{align*}
	F^{1,0} 	&\,=\, dA^{1,0} \,=\, d\star d(\, \alpha\, dz \w dt\,) \\[0.2cm]
		&\,=\, \frac{\partial^{2}\alpha}{\partial t\,\partial x}\,e^{0}\vee e^{2} -\frac{\partial^{2}\alpha}{\partial t\,\partial y}\,e^{0} \vee e^{3} + \frac{\partial^{2}\alpha}{\partial x\,\partial z}\,e^{1}\vee e^{2} - \frac{\partial^{2}\alpha}{\partial x\,\partial y}\,e^{1}\vee e^{3} - \left(\frac{\partial^{2}\alpha}{\partial y^{2}}+\frac{\partial^{2}\alpha}{\partial z^{2}}\right)\,e^{2}\vee e^{3} 
\end{align*}
for any complex $0$-form $\alpha\in\Gamma\Lambda^{0}\M$. A direct computation now verifies that if $\alpha$ is the complex pulse solution (\ref{alp}) satisfying 
\begin{align*}
	\Box\,\alpha \,=\, \delta d\alpha \,=\, \frac{\partial^{2}\alpha}{\partial t^{2}} - \frac{\partial^{2}\alpha}{\partial x^{2}} - \frac{\partial^{2}\alpha}{\partial y^{2}} - \frac{\partial^{2}\alpha}{\partial z^{2}} \,=\, 0
\end{align*}
then the spinor field $\Psi^{1,0}=\Psi^{1,0}\vee \proj_{1}\equiv F^{1,0}\vee \proj_{1}\in\Gamma\mathcal{I}_{L}\proj\M$ satisfies the massless spinor field equation: 
\begin{align}\label{RWT_SSPSI}
	\slashed{S}\Psi^{1,0} \,=\, 0. 
\end{align}
Similarly, in the cylindrical polar chart we write $\Phi^{1,0}\equiv \,=\, F^{1,0} \vee \PROJ_{1} \,=\, \Phi^{1,0} \vee \PROJ_{1}  \in \Gamma\mathcal{I}_{L}\PROJ_{1}\M$ and also find 
\begin{align}\label{RWT_SSPHI}
	\slashed{S}\Phi^{1,0} \,=\, 0. 
\end{align}
This follows since $\Phi^{1,0}$ and $\Psi^{1,0}$ are (differentiable) spinor fields equivalent under the action of elements of the Clifford group that relate the generators of the ideals in different coordinate charts. To appreciate the inner consistency of the above definitions and the significance of these results, we have for any spinor field $\Xi$:
\begin{align*}
	\slashed{S}\,\Xi \,\equiv\, e^{a} \vee S_{X_{a}}\Xi.
\end{align*}
Then for {\it any} primitive idempotent $\proj$ and $2$-form $\pmb{F}$, with $\Xi=\pmb{F}\vee\proj$:
\begin{align*}
	e^{a} \vee S_{X_{a}}(\pmb{F} \vee \proj) &\,=\, e^{a} \vee (S_{X_{a}}\pmb{F} \vee \proj + \pmb{F} \vee S_{X_{a}}\proj) \\[0.2cm]
		&\,=\, e^{a} \vee \nabla_{X_{a}}\pmb{F} \vee \proj + e^{a} \vee \pmb{F} \vee S_{X_{a}}\proj \\[0.2cm]
		&\,=\, e^{a} \w \nabla_{X_{a}}\pmb{F} \vee \proj + i_{\wt{e^{a}}}\nabla_{X_{a}}\pmb{F} \vee \proj + e^{a} \vee \pmb{F} \vee S_{X_{a}}\proj
\end{align*}
using the spinor Leibniz rule and (\ref{RWT_HASH1}). Thus
\begin{align}\label{RWT_NB_08}
	\slashed{S}(\pmb{F} \vee \proj)  &\,=\, d\pmb{F} \vee \proj -\delta \pmb{F} \vee \proj + e^{a} \vee \pmb{F} \vee S_{X_{a}}\proj.
\end{align}
Hence if $\pmb{F}$ is {\it any} closed and co-closed $2$-form, the first two terms on the right-hand side of (\ref{RWT_NB_08}) vanish. Furthermore, in the adapted global $\nabla$-parallel Cartesian chart, $S_{X_{a}}\proj=0$ ($a=0,1,2,3$) and one has the result (\ref{RWT_SSPSI}) with $\Xi\equiv\Psi^{1,0}$. The covariance of the spinor covariant derivative under ideal transformations induced by coordinate transformations is manifest by the vanishing of all terms on the right-hand side of (\ref{RWT_NB_08}) in the above cylindrically polar Minkowski chart with $\proj=\PROJ$. This again yields the result (\ref{RWT_SSPHI}) with $\Xi\equiv\Psi^{1,0}$. However, in general not all polar primitives $\PROJ'$ may generate a solution of the form $\pmb{F} \vee \PROJ'$ for closed and co-closed $\pmb{F}$ since not all polar primitives $\PROJ'$ satisfy $S_{X_{a}}\PROJ'=0$ with $J=\xi^{*}$ for all basis elements in the span of $\{X_{a}\}$. Nevertheless, among the solutions to (\ref{RWT_SSPHI}) we may construct propagating, non-singular, spinor pulse-like chiral solutions based upon the pre-potential (\ref{CYL_ALP}):
\begin{align}\label{SPINOR_alp}
   \alpha(t,\rho,z) \,=\, \frac{\W^{2}}{\rho^{2} + [\,a + i(z-t)\,]\,[\,b-i( z+t )\,]  }
\end{align}
and any complex $2$-form $\Pi^{\nu,\chi}$ given by (\ref{SPINOR_PI}). \\

In sections~\ref{sect:EM} and \ref{sect:EIN} we have indicated how to generate new vacuum propagating multi-chiral pulse solutions to the source-free Maxwell, Bopp-Podolsky-Land\'{e} and linearised Einstein gravitational field equations in a Minkowski spacetime background by taking suitable derivatives of the {\it tensor fields} constructed from solutions based upon certain pre-potentials, with respect to Killing vector fields that generate the rotation groups $O(2)$. Although the operator $\Lie_{X}$ is well-defined for any vector field $X\in\Gamma T\M$, it will only act as a derivation on a Clifford product of forms if $X$ is a (conformal) Killing vector field $K$ of $\GG$. In that case, we have for the Levi-Civita covariant derivative $\nabla$ the identity:
\begin{align*}
	\Lie_{K}\Omega \,=\, \nabla_{K}\Omega + \left[\frac{1}{4}d\wt{K},\Omega\right] \qquad\text{for all   }\Omega \in \Gamma \textrm{C}_{3,1}(V,\GG)\M
\end{align*} 
in terms of the Clifford commutator (\ref{CliffCOMM}). If $K_{1},K_{2}$ are any (conformal) Killing vector fields with {\it Lie bracket} $[K_{1},K_{2}]\equiv \Lie_{K_{1}}K_{2}$ one has the identity:
\begin{align*}
	\left[\Lie_{K_{1}}, \Lie_{K_{2}}\right] \,=\, \Lie_{[K_{1},K_{2}]}
\end{align*}
acting on {\it tensor} fields. This motivates the definition of a {\it spinor Lie derivative} operator $\sLie_{K}$ acting on any spinor field $\Psi\in\Gamma\mathcal{I}_{L}\proj\M$ where $K$ is any (conformal) Killing vector field on $\M$:
\begin{align*}
	\sLie_{K} \Psi \,\equiv\, S_{K}\Psi  + \frac{1}{4}d\wt{K} \vee \Psi,
\end{align*}
since the right-hand side stays within the ideal generated by $\proj$ and in terms of Clifford commutators:
\begin{align*}
	\left[\sLie_{K_{1}}, \sLie_{K_{2}}\right] \,=\, \sLie_{[K_{1},K_{2}]}
\end{align*}
for all (conformal) Killing vector fields $K_{1},K_{2}$. \\

As noted above, the {\it chiral} properties of tensor fields constructed from the pre-potential $\alpha$ and chiral $2$-form $\Pi^{\nu,\chi}$ are defined with respect to the Killing vector field $\partial/\partial\phi$ and the tensor operators $\nabla_{\partial/\partial\phi}$ or $\Lie_{\partial/\partial\phi}$. The spinor fields $\Psi^{1,0}$ and $\Phi^{1,0}$ above are defined in terms of projector idempotents adapted to either the Cartesian or cylindrical polar chart in Minkowski spacetime. These projector sets are special since they are both associated with the differential operator $\Lie_{\partial/\partial\phi}$ as well as $\slashed{S}$ (when $\Box\,\alpha=0$):
\begin{align*}
		\Lie_{\partial/\partial\phi}\proj_{i} \,=\, \Lie_{\partial/\partial\phi}\PROJ_{i} \,=\, \Lie_{\partial/\partial\phi}\wh{\PROJ}_{i} \,=\, 0 \qquad (i=1,2,3,4).
\end{align*}
This then implies that any spinor solution generated by a primitive $\proj$ of the form:
\begin{align}\label{RWT_eee}
	\Xi^{\nu,\chi} \,=\, F^{\nu,\chi}\vee \proj 
\end{align}
with $\Lie_{\partial/\partial\phi}\proj=0$ satisfies 
\begin{align}\label{RWT_labela}
	\slashed{S}\,\Xi^{\nu,\chi} \,=\, 0.
\end{align}
It is notable that using projectors adapted to different coordinate systems the spinor Lie derivative plays no role in this construction. \\

Furthermore, by analogy with the construction of multi-chiral Maxwell solutions, it follows that for all chiral solutions $F^{\nu,\chi}$ (\ref{Fnuchi}) we can construct a family of {\it multi-chiral spinor fields} defined by:
\begin{align*}
	\Xi^{\nu,r\pm\chi} \,\equiv\, F^{\nu,r\pm\chi}\vee \proj \,=\, \left(\Lie_{K_{\pm}}^{r}\,F^{\nu,\chi}\right) \vee \proj \quad\text{for all $\proj$ such that $\Lie_{\partial/\partial\phi}\proj=0$}
\end{align*}
where $K_{\pm}$ is the complex Killing vector (\ref{Kpm}) generating translations in the transverse $(x,y)$-plane. Then
\begin{align}\label{RWT_labelb}
	\frac{1}{i}\,\Lie_{\partial/\partial \phi}\,\Xi^{\nu,r\pm\chi} \,=\, (r\pm\chi)\,\Xi^{\nu,r\pm\chi} \qquad (r=0,1,2,\ldots)
\end{align}
(i.e. the spinor $\Xi^{\nu,r\pm\chi}$ has chirality $r\pm\chi$). For projectors that also satisfy: 
\begin{align}\label{eAFSP}
	e^{a} \vee F^{\nu,r\pm\chi} \vee S_{X_{a}}\proj \,=\, 0 \qquad (r=0,1,2,\ldots)
\end{align}
it follows from (\ref{RWT_NB_08}) that such fields are zero-mass Dirac solutions:
\begin{align}\label{XiDirac}
	\slashed{S}\,\Xi^{\nu,r\pm\chi} \,=\, 0.
\end{align}
For $\proj$ belonging to $\{\proj_{i}\}$ and $\{\PROJ_{i}\}$ ($i=1,2,3,4$) we have therefore shown solutions to (\ref{eAFSP}), and hence (\ref{XiDirac}), do exist. Amongst such solutions are those describing a $3$-parameter family of propagating, non-stationary, non-singular, pulse-like spinor fields, spatially bounded in all three dimensions.\\

In summary, if $\proj\in\Gamma\mathcal{I}_{L}\proj\M$ is an $S$-parallel Clifford idempotent satisfying 
$\Lie_{\partial/\partial\phi}\proj=0$ where $\partial/\partial\phi$ is the real Minkowski Killing vector field given in (\ref{DEF_Kphi}), $\alpha$ any complex scalar field satisfying $\Box\,\alpha=0$ and $\Pi^{\nu,\chi}$ a $\nabla$-parallel $2$-form in Minkowski spacetime satisfying (\ref{SPINOR_PI}), then the spinor field $\Xi^{\nu,\chi}\equiv F^{\nu,\chi} \vee \proj$ with $F^{\nu,\chi}\equiv dA^{\nu,\chi}$, $A^{\nu,\chi}\equiv \star d(\,\alpha\Pi^{\nu,\chi}\,)$ will satisfy the Dirac spinor equation (\ref{RWT_labela}). All such solutions have chirality $\chi=\{-1,0,1\}$. Particular solutions with $\alpha$ given by (\ref{SPINOR_alp}) generate a $3$-parameter family of non-singular propagating spinor pulses with chirality $\chi$. Furthermore, in general, spinor fields $\Xi^{\nu,\chi\pm r}$ with chirality $\chi\pm r$ ($r=0,1,2,\ldots$) can be generated from $F^{\nu,\chi}$ as indicated in (\ref{RWT_labelb}) by Lie differentiation with respect to the complex Killing vector field $K_{\pm}$ (\ref{Kpm}). \\

\section{Concluding Remarks}\label{sect:CONC}
In this article we have illustrated a technique for analysing a class of linear partial differential equations in terms of non-separable solutions $\alpha$ to the complex scalar Lorentz-covariant Laplace equation $\Box\,\alpha=0$ and a set of complex, anti-symmetric, rank-two eigen-tensors of a chiral operator in Minkowski spacetime. The technique has been applied to the Maxwell, Bopp-Land\'{e}-Podolsky, linearised Einstein and massless electrically-neutral Dirac field systems in source-free spacetime domains. Particular solution families have been constructed for each system that exhibit axially symmetric, dispersive, propagating solutions bounded in all three spatial dimensions of these domains. \\

Since neutrinos are no longer considered to be massless fundamental particles, the massless Dirac spinor pulses derived here lie outside the Standard Model of particle physics. In principle, such chiral spinor beams could affect the time-like geodesic structure of spacetime via their coupling to gravitation through their stress-energy-momentum tensor and hence influence the geodesic motion of massive test particles. This is analogous to our illustration of how the chiral Maxwell pulses influence the motion of electrically-charged test particles via their coupling through the covariant Lorentz force and to the chiral gravitational pulses that initiate astrophysical jets. The generation of fields with chiral characteristics is a common feature of all solutions constructed with the use of our pre-potentials. As we have shown, these features can be used to model phenomenology on vastly different physical scales. In particular, they may also offer scope for exploring observable properties of pulsars and the formation of spiral galaxies, as well as astrophysical jets. \\

A further direction for applications is to extend the technique to linear spinor-tensor field equations (for example, the Rarita-Schwinger field equation for spin-$3/2$ particles). This may offer phenomenological models for systems that exhibit linearised supergravity or supersymmetry (in any spacetime dimension). In this context, we comment that both tensor and spinor field systems arise naturally as different sections of a Clifford bundle over spacetime. Furthermore, the even sub-algebra of the Clifford algebra admits spinor representations (Weyl spinors) and spinor field equations that are also amenable to analysis in terms of chiral pre-potentials. \\

Finally we note that the chirality of classical free-space propagating solutions should not be confused with the notion of ``helicity''. In particular, the helicity of free-space photons, regarded as the quanta associated with free-space plane waves, is restricted to $\pm 1$ and is defined in terms of source-free Maxwell separable-mode solutions. Such modes have no longitudinal polarisation in the direction of propagation unlike propagating free-space chiral modes. Whether it is useful to regard any of the particular propagating chiral solutions described in this paper as a Fourier-superposition of quantised fundamental particle (separable) modes, we shall leave to the reader. \\

\section*{Acknowledgements}
It is with great pleasure for the authors to pay their respects to Tekin Dereli on his $70$th birthday Festschrift celebration. He and RWT have for many years been close research colleagues and family friends. One of the immeasurable benefits arising from scientific collaborations abroad can be discovering the history and cultural heritage of other countries. Besides their mutual enthusiasm for physics, RWT will be forever grateful to Tekin for revealing to him both the history and geography of his beautiful country. \\

The authors RWT and TJW are also grateful to STFC (ST/G008248/1) and EPSRC (EP/J018171/1) for support, and acknowledge useful discussions with colleagues in the ALPHA-X project and Cockcroft Institute. \\

\bibliographystyle{unsrt}
\bibliography{PREPOTS}

\end{document}

%% file: NEW_TEKIN_FIG_01.pdf_tex
\begingroup%
  \makeatletter%
  \providecommand\color[2][]{%
    \errmessage{(Inkscape) Color is used for the text in Inkscape, but the package 'color.sty' is not loaded}%
    \renewcommand\color[2][]{}%
  }%
  \providecommand\transparent[1]{%
    \errmessage{(Inkscape) Transparency is used (non-zero) for the text in Inkscape, but the package 'transparent.sty' is not loaded}%
    \renewcommand\transparent[1]{}%
  }%
  \providecommand\rotatebox[2]{#2}%
  \newcommand*\fsize{\dimexpr\f@size pt\relax}%
  \newcommand*\lineheight[1]{\fontsize{\fsize}{#1\fsize}\selectfont}%
  \ifx\svgwidth\undefined%
    \setlength{\unitlength}{1036.5bp}%
    \ifx\svgscale\undefined%
      \relax%
    \else%
      \setlength{\unitlength}{\unitlength * \real{\svgscale}}%
    \fi%
  \else%
    \setlength{\unitlength}{\svgwidth}%
  \fi%
  \global\let\svgwidth\undefined%
  \global\let\svgscale\undefined%
  \makeatother%
  \begin{picture}(1,0.36034732)%
    \lineheight{1}%
    \setlength\tabcolsep{0pt}%
    \put(0,0){\includegraphics[width=\unitlength,page=1]{NEW_TEKIN_FIG_01.pdf}}%
    \put(0.12161819,0.32134092){\color[rgb]{0,0,0}\makebox(0,0)[lt]{\lineheight{1.25}\smash{\begin{tabular}[t]{l}{\large (CM,$1$)}\end{tabular}}}}%
    \put(0.44352322,0.32134092){\color[rgb]{0,0,0}\makebox(0,0)[lt]{\lineheight{1.25}\smash{\begin{tabular}[t]{l}{\large (CM,$0$)}\end{tabular}}}}%
    \put(0.75034453,0.32134092){\color[rgb]{0,0,0}\makebox(0,0)[lt]{\lineheight{1.25}\smash{\begin{tabular}[t]{l}{\large (CM,$-1$)}\end{tabular}}}}%
    \put(0.03631883,0.16568664){\color[rgb]{0,0,0}\makebox(0,0)[lt]{\lineheight{1.25}\smash{\begin{tabular}[t]{l}$x$\end{tabular}}}}%
    \put(0.35347749,0.16568641){\color[rgb]{0,0,0}\makebox(0,0)[lt]{\lineheight{1.25}\smash{\begin{tabular}[t]{l}$x$\end{tabular}}}}%
    \put(0.67214512,0.16568641){\color[rgb]{0,0,0}\makebox(0,0)[lt]{\lineheight{1.25}\smash{\begin{tabular}[t]{l}$x$\end{tabular}}}}%
    \put(0.13607951,0.0405978){\color[rgb]{0,0,0}\makebox(0,0)[lt]{\lineheight{1.25}\smash{\begin{tabular}[t]{l}$y$\end{tabular}}}}%
    \put(0.45831614,0.04059826){\color[rgb]{0,0,0}\makebox(0,0)[lt]{\lineheight{1.25}\smash{\begin{tabular}[t]{l}$y$\end{tabular}}}}%
    \put(0.77102741,0.04059826){\color[rgb]{0,0,0}\makebox(0,0)[lt]{\lineheight{1.25}\smash{\begin{tabular}[t]{l}$y$\end{tabular}}}}%
    \put(0.26606567,0.05264564){\color[rgb]{0,0,0}\makebox(0,0)[lt]{\lineheight{1.25}\smash{\begin{tabular}[t]{l}$z$\end{tabular}}}}%
    \put(0.58814404,0.05264517){\color[rgb]{0,0,0}\makebox(0,0)[lt]{\lineheight{1.25}\smash{\begin{tabular}[t]{l}$z$\end{tabular}}}}%
    \put(0.90319358,0.04975081){\color[rgb]{0,0,0}\makebox(0,0)[lt]{\lineheight{1.25}\smash{\begin{tabular}[t]{l}$z$\end{tabular}}}}%
  \end{picture}%
\endgroup%

%% file: NEW_TEKIN_FIG_00.pdf_tex
\begingroup%
  \makeatletter%
  \providecommand\color[2][]{%
    \errmessage{(Inkscape) Color is used for the text in Inkscape, but the package 'color.sty' is not loaded}%
    \renewcommand\color[2][]{}%
  }%
  \providecommand\transparent[1]{%
    \errmessage{(Inkscape) Transparency is used (non-zero) for the text in Inkscape, but the package 'transparent.sty' is not loaded}%
    \renewcommand\transparent[1]{}%
  }%
  \providecommand\rotatebox[2]{#2}%
  \newcommand*\fsize{\dimexpr\f@size pt\relax}%
  \newcommand*\lineheight[1]{\fontsize{\fsize}{#1\fsize}\selectfont}%
  \ifx\svgwidth\undefined%
    \setlength{\unitlength}{689.25bp}%
    \ifx\svgscale\undefined%
      \relax%
    \else%
      \setlength{\unitlength}{\unitlength * \real{\svgscale}}%
    \fi%
  \else%
    \setlength{\unitlength}{\svgwidth}%
  \fi%
  \global\let\svgwidth\undefined%
  \global\let\svgscale\undefined%
  \makeatother%
  \begin{picture}(1,0.80195865)%
    \lineheight{1}%
    \setlength\tabcolsep{0pt}%
    \put(0,0){\includegraphics[width=\unitlength,page=1]{NEW_TEKIN_FIG_00.pdf}}%
    \put(0.27978781,0.13615343){\color[rgb]{0,0,0}\makebox(0,0)[lt]{\begin{minipage}{0.24537541\unitlength}\raggedright \end{minipage}}}%
    \put(0.01116951,0.12456036){\color[rgb]{0,0,0}\rotatebox{91.24282}{\makebox(0,0)[lt]{\begin{minipage}{0.49673558\unitlength}\raggedright \end{minipage}}}}%
    \put(0.32916213,0.12744831){\color[rgb]{0,0,0}\makebox(0,0)[lt]{\begin{minipage}{0.10922198\unitlength}\raggedright $x$\end{minipage}}}%
    \put(0.80600091,0.23912924){\color[rgb]{0,0,0}\makebox(0,0)[lt]{\begin{minipage}{0.10922198\unitlength}\raggedright $y$\end{minipage}}}%
    \put(0.00837483,0.24){\color[rgb]{0,0,0}\rotatebox{89.033909}{\makebox(0,0)[lt]{\begin{minipage}{0.34711643\unitlength}\raggedright Power profile, P\end{minipage}}}}%
  \end{picture}%
\endgroup%